\documentclass[review,authoryear,3p,times,10pt]{elsarticle}

\usepackage{multirow,setspace,times,amssymb,amsmath,graphicx,color,rotating,subfigure,url}
\usepackage{lineno,color}
\usepackage{natbib}
\usepackage{booktabs}%
\usepackage{longtable}%
\usepackage{rotating}
\usepackage{lscape}
\usepackage{pdflscape}
\usepackage{ulem}
\usepackage{threeparttable}
\usepackage{bm}

\graphicspath{{Figures/}}
\usepackage[table]{xcolor}
\usepackage{tabularx}
\usepackage{graphicx} 
\usepackage{epstopdf}
\usepackage{mathrsfs}
\usepackage{makecell}
\usepackage[bookmarks=true,colorlinks,linkcolor=blue,anchorcolor=blue,citecolor=blue,unicode]{hyperref}
\usepackage{bookmark}

\usepackage{todonotes}
\usepackage{ragged2e}
\usepackage[font=small]{caption}

\hypersetup{CJKbookmarks=true}%
\journal{Journal of Futures Markets}

\begin{document}

\begin{frontmatter}
\title{Uncovering the Sino-US dynamic risk spillovers effects: Evidence from agricultural futures markets}

\author[SB]{Han-Yu Zhu}
 \author[WHUT1,WHUT2]{Peng-Fei Dai\corref{CorAuth}}
 \ead{pfdai@ecust.edu.cn}
 \author[SB,RCE,DM]{Wei-Xing Zhou\corref{CorAuth}}
\ead{wxzhou@ecust.edu.cn}
 \cortext[CorAuth]{Corresponding author.} 

\address[SB]{School of Business, East China University of Science and Technology, Shanghai 200237, China}

\begin{abstract}
  Agricultural products play a critical role in human development. With economic globalization and the financialization of agricultural products continuing to advance, the interconnections between different agricultural futures have become closer. We utilize a TVP-VAR-DY model combined with the quantile method to measure the risk spillover between 11 agricultural futures on the futures exchanges of US and China from July 9, 2014, to December 31, 2022. This study yielded several significant findings. Firstly, CBOT corn, soybean, and wheat were identified as the primary risk transmitters, with DCE corn and soybean as the main risk receivers. Secondly, sudden events or increased economic uncertainty can increase the overall risk spillovers. Thirdly, there is an aggregation of risk spillovers amongst agricultural futures based on the dynamic directional spillover results. Lastly, the central agricultural futures under the conditional mean are CBOT corn and soybean, while CZCE hard wheat and long-grained rice are the two risk spillover centers in extreme cases, as per the results of the spillover network and minimum spanning tree. Based on these results, decision-makers are advised to safeguard against the price risk of agricultural futures under sudden economic events, and investors can utilize the results to construct a superior investment portfolio by taking different agricultural product futures as risk-leading indicators according to various situations.
\end{abstract}


\end{frontmatter}


\section{Introduction}

The production and trade of agricultural products are closely related to people's daily lives, especially when it comes to food, but food crises are frequent. From 2000 to 2004, there were about 850 million people affected by hunger in the world; in 2010, the number of people affected by hunger in the world reached 1 billion; and in April 2022, due to the COVID-19 pandemic and the Russia-Ukraine conflict, the United Nations World Food Program announced that humanity may face ``the largest food crisis since World War II''. The food crisis is usually characterized by excessive price increases. Therefore, this paper focuses on the price risk of agricultural products.

The mitigation of agricultural product price fluctuations through hedging has contributed to the emergence and evolution of agricultural futures contracts, marking their pioneering role in global futures markets. Despite the growth of futures markets and the expansion of trading options, agricultural futures remain a critical component, making indispensable contributions to the broader futures and financial landscapes. Presently, an extensive array of over seventy agricultural products is actively traded across global futures markets, categorized into four main segments: staple foods, cash crops, livestock, and forest products. Remarkably, agricultural futures have witnessed substantial trading activity, surpassing even the trading volume of agricultural spot markets, as indicated by the American Futures Association's data, which reported 1.815 billion lots of global agricultural futures traded in 2021, accounting for 9\% of the total global futures trading volume.

Futures markets serve a fundamental role in price discovery and hedging, enabling the establishment of reference prices for the spot market and facilitating price exploration within it. When it comes to agricultural products, price discovery exhibits a distinct unidirectional flow from the futures market to the spot market. This dynamic imparts the potential for the transmission of price-related risks into the spot market. Moreover, since 2004, the commodity market has witnessed a notable influx of financial capital, leading to the financialization of commodities, including agricultural products. This integration of financial instruments has intensified the connectedness of price fluctuations across various agricultural commodity futures markets. Furthermore, changes in the import and export trade volume and structure of agricultural spot markets contribute to heightened price risk in agricultural futures. Taking corn as an example, global imports and exports of corn reached 1,787,579,000 tons and 200,443,000 tons in 2021, an increase of 27.45\% and 65.94\% from 2015, respectively. The intricate dynamics of price volatility and the transmission of risk events are further compounded by the continuous financialization of agricultural products, resulting in increasingly complex mechanisms.

The phenomenon of risk transmission from one market to other markets is called ``risk spillover''. In the context of economic globalization and financial integration, the degree of correlation between different types of agricultural futures in different regions is increasing. As a result, the volatility of the entire agricultural futures market increases when a certain shock occurs, and  they are more likely to interact with each other, leading to sharp increases and decreases in the same direction and increasing the risk of the entire market. The linkage  between risk transmission has become more apparent since the beginning of the 21st century. For example, from 2006 to 2008, due to the impact of extreme weather, agricultural production decreased, and coupled with the rising cost of crude oil prices, agricultural futures prices generally increased. CBOT corn prices reached a maximum increase of 274\% in June 2008, and CBOT wheat prices reached a maximum increase of 323\% in February 2008, which had a large impact on the entire agricultural futures market. The rising prices of agricultural products in turn had a serious impact on developing countries, with a number of import-dependent countries experiencing ``food riots'' one after another, resulting in casualties.

Various methods are employed to investigate risk spillover in academic research, including quantile regression, multivariate GARCH, copula models, and vector autoregression. For the present study, which focuses on assessing both average price risk spillover and extreme price risk spillover among agricultural futures, it is essential to utilize a model that adequately captures the measurement requirements concerning direction, intensity, and the time-varying nature of risk spillover due to the normalization of shocks. In this study, we chose to use the TVP-VAR-DY model, supplemented by the quantile VAR, to construct the DY spillover index. This index serves to measure the price risk spillover of agricultural futures by examining the time-varying characteristics of price risk. Moreover, the model enables the evaluation of the intensity and direction of risk spillovers from one market to all other markets, as well as pairwise risk spillovers between any two markets, thus providing a comprehensive measurement of price risk at the aggregate systemic level.

With the continuous development of economic globalization, the interdependence among nations has intensified, leading to accelerated and amplified transmission of agricultural risk shocks across countries. This study investigates the dynamics of risk spillovers from the occurrence of agricultural shocks in one country to others, with a particular focus on staple food futures between the United States and China. In the spot market, the United States holds the position of the world's largest producer of soybeans and corn and the largest exporter of food, while China is the largest producer of wheat and rice. Given its population, China is by far the largest importer of food, resulting in close import-export transactions for these agricultural commodities. Furthermore, China's agricultural futures market has demonstrated remarkable growth, with its top ten agricultural futures contributing significantly to the global agricultural futures turnover in 2022, while the growth rate of the more established US agricultural futures market has shown signs of deceleration. Given the evolving landscape of futures markets, this paper aims to empirically re-evaluate the nature of risk spillovers between US and Chinese agricultural futures. By utilizing data from 11 staple food futures listed in China and the United States, spanning the period from July 9, 2014, to December 31, 2022, we provide insights into the dynamics of risk transmission and the implications for market participants and policymakers.

The rest of this article is organized as follows. Section~\ref{S1:LitRev} reviews the literature. Section~\ref{S1:Methodology} describes the data set and the TVP-VAR-DY model. Section~\ref{S1:Data} describes the empirical dataset and presents the statistical description. Section~\ref{S1:EmpAnal} provides an empirical examination of the risk spillovers between agricultural futures from both conditional mean and quantile perspectives, offering visual analysis. Section~\ref{S1:Conclude} contains some conclusions and implications. 

\section{Literature review}
\label{S1:LitRev}

\subsection{Agricultural futures market}


Price discovery is one of the basic functions of futures, and whether price discovery can be effective and how efficient the market is has been a hot issue of concern for scholars. On the theoretical side, scholars have proposed various theories to explain the price discovery mechanism of futures, including \cite{Michael-Brennan-1958-TAmericanEconomicReview}'s cost-of-ownership theory and \cite{Cootner-1960-JPE}'s normal premium/discount theory. Empirically, \cite{FJ-McKenzie-Holt-2002-ApplEcon} examine market efficiency and unbiasedness in four agricultural futures markets and find that while there are some market biases and inefficiencies in the short run, each market is unbiased and efficient in the long run. \cite{FJ-Li-Xiong-2021-JAsianEcon} study the price discovery function of China's agricultural futures market and obtain the conclusion that most agricultural futures are effective in price discovery. In addition, with market-oriented policy changes, most futures markets have some enhancement of price discovery function. \cite{FJ-Ke-Li-McKenzie-Liu-2019-Sustainability} apply the CoVaR method to explore risk transmission between the Chinese and US agricultural futures markets for soybeans, corn, and sugar, confirming the dominant pricing role of the US agricultural futures market, while acknowledging the growing role of the Chinese market in price discovery.

Agricultural futures prices have their own price characteristics and exhibit sudden and unexpected price jumps \citep{FJ-Koekebakker-Lien-2004-AmJAgrEcon}. The volatility of agricultural futures prices is not conducive to their functioning, and there are many factors that can contribute to the volatility of agricultural futures prices. Early studies were mainly explained in terms of supply and demand imbalances, but as the number of speculators in agricultural futures trading increased, more studies conducted analyses of financialization factors. \cite{FJ-AitYoucef-2019-EconModel} finds that the price, volatility, and correlation of agricultural futures increased with financialization and that agricultural futures market volatility was more influenced by the stock market during the financial crisis, which is consistent with \cite{FJ-Basak-Pavlova-2016-JFinanc}. Using quantile regressions, \cite{FJ-Bianchi-Fan-Todorova-2020-IntRevFinancAnal} find that agricultural futures price volatility is significantly influenced by financialization, although at a less significant level than energy commodity futures.

The intensifying global economic integration has fostered intricate interconnections among agricultural futures across diverse regions, as well as their interplay with various other commodity futures. \cite{Hernandez-Ibarra-Trupkin-2014-ERAE} use a multivariate GARCH model to study the volatility dynamics among the major global agricultural futures markets in the United States, Europe, and Asia, and conclude that the agricultural futures markets are highly interconnected and have a growing trend over time. \cite{FJ-Jia-Wang-Tu-Li-2016-PhysicaA} explore the excess-lag relationship between the US and Chinese agricultural futures markets and show that Chinese soybean, corn, and wheat futures price volatility were all ahead of the US after 2014. Research on the linkages between agricultural futures and other commodity futures has centered on the crude oil market and the agricultural futures market. The general agreement in the studies is that the overall linkage between the crude oil market and the agricultural futures market has become stronger since the 2006-2008 food crisis \citep{FJ-Kristoufek-Janda-Zilberman-2012-EnergyEcon}. Subsequent studies focus on market segments, such as \cite{FJ-Luo-Ji-2018-EnergyEcon} verify the spillover effect of the volatility of the US crude oil market on the Chinese agricultural market and the existence of leverage effects of this volatility transmission. 

Some scholars have studied the linkages and risk transfers between different agricultural futures. \cite{FJ-Gardebroek-Hernandez-Robles-2016-AgricEcon} conducts an investigation into the conditional correlation and temporal dynamics of volatility transmission within the context of CBOT-listed corn, wheat, and soybean futures price returns, revealing that the transmission of volatility is significantly influenced by the wheat and corn. \cite{FJ-Zivkov-Kuzman-Subic-2020-AgricEcon} use Bayesian quantile to study price volatility of CBOT-listed corn, wheat, soybean, and rice futures and find that rice was subject to the smallest volatility shocks from all other agricultural futures and that risk shocks were asymmetric, with stronger volatility spillover effects during periods of market turmoil. The literature mainly focuses on CBOT-listed corn, wheat, soybeans, and other staple food futures.

\subsection{Risk spillover models}

Research on the risk spillover effect is often practiced in different markets. The main research can be broadly divided into risk spillovers between different types of markets and risk spillovers between different subjects or products in the same type of market. The former includes risk spillovers between the stock market and the foreign exchange market, risk spillovers between the stock market and the crude oil market, and so on \citep{FJ-Chen-Wen-Li-Yin-Zhao-2022-EnergyEcon, FJ-Du-He-2015-EnergyEcon}, and in the latter, risk spillovers between banks in different regions and risk spillovers in stock markets are included \citep{FJ-Apostolakis-Floros-Giannellis-2022-IntRevEconFinanc,FJ-Su-2020-NAmEconFinanc}.

Earlier studies on risk spillovers were investigated based on traditional methods like the quantile model. The quantile model is good for dealing with the tail dependence problem with asymmetric left and right tails. \cite{FJ-Iqbal-Bouri-Grebinevych-Roubaud-MISSING-AnnOperRes} use the quantile connectivity measure and find that realized volatility shocks circulate more intensely during extreme events relative to normal periods. The quantile model has the defect of not applying to the study of the clustering effect of yield volatility, and the multivariate GARCH model as a new model can make up for its defect. Applying this model, \cite{FJ-Mensi-Hammoudeh-Nguyen-Yoon-2014-EnergyEcon} obtain evidence of a significant link between energy and grain markets. \cite{FJ-Kang-McIver-Yoon-2017-EnergyEcon} use a multivariate GARCH model in conjunction with a spillover index model and show that gold and silver are information transmitters to other commodity futures markets during times of financial stress, while the remaining four investigated energy and agricultural futures are recipients of risk spillovers. However, the GARCH model has certain requirements for the division and relationship of the variables. The Copula model does not suffer from the disadvantages of the first two models. \cite{FJ-Ji-Bouri-Roubaud-Shahzad-2018-EnergyEcon} use a time-varying Copula model with switching dependence to study the conditional dependence between energy and agricultural markets and find that the lower-tailed dependence is much stronger in the bearish state than in the bullish state. \cite{FJ-Tiwari-Boachie-Suleman-Gupta-2021-Energy} use a Copula-based model for their study and find that corn, oats, and wheat have the ability to hedge against declining oil returns due to geopolitical turmoil.

From the perspective of risk metrics, VaR is a traditional risk metric; however, this metric and the corresponding model can only measure the risk of a variable itself, not the risk spillover effect between different subvariables within the variable. Therefore, based on the VaR metrics, \cite{FJ-Adrian-Brunnermeier-2016-AmEconRev} propose the CoVaR approach, providing new ideas for risk spillover research. CoVaR can be calculated in combination with the three models mentioned above. But CoVaR is generally used to measure volatility risk spillover.

For mean risk spillover, scholars use vector autoregressive VAR for risk spillover effect analysis. \cite{FJ-Yip-Brooks-Do-Nguyen-2020-IntRevFinancAnal} find that the net volatility spillover effect of crude oil on all agricultural commodities tends to decrease when crude oil remains in a low volatility state by applying the VAR model. \cite{FJ-Barbaglia-Croux-Wilms-2020-EnergyEcon} use an optimized t-LASSO vector autoregressive model to study the volatility spillovers among energy, biofuel, and agricultural markets, confirming that the volatility spillovers in agricultural and energy markets are linked by biofuel markets. However, the variable ordering in the traditional VAR process may lead to different results.

Based on the VAR model, \cite{FJ-Diebold-Yilmaz-2009-EconJ,FJ-Diebold-Yilmaz-2012-IntJForecast} innovatively propose the method of spillover index. They optimize the variance decomposition in the original model to the generalized forecast variance decomposition and construct the total spillover index and directional spillover index according to the matrix obtained from the forecast error variance decomposition. Meanwhile, they combine the rolling window method to analyze the spillover index from a dynamic perspective. \cite{FJ-Adeleke-Awodumi-2022-JApplEcon} use DY spillover indices to analyze the time-domain volatility correlations in energy, agricultural raw materials, and food markets. Dynamic time-domain studies of the DY spillover index require the application of a rolling window, but the rolling window needs to be set in size, which is subjective and carries the risk of losing observations. Therefore, \cite{Antonakakis-Chatziantoniou-Gabauer-2020-RFM} improve and innovate the DY index based on the TVP-VAR model to  investigate the dynamic risk premium index measures for the four most traded foreign exchange rates.

The main advantages of the TVP-VAR-DY model are that the dynamic risk spillover index measure can be performed without a rolling window to avoid losing observations, and that the model is less sensitive to outliers due to the multivariate Kalman filtering approach. The TVP-VAR-DY model is currently applied to the stock market, the energy market, and the foreign exchange market, focusing on the variation of risk premiums over time \citep{FJ-Cheng-Deng-Liang-Cao-2023-ResourPolicy,FJ-Duan-Xiao-Ren-TaghizadehHesary-Duan-2023-ResourPolicy,FJ-Tiwari-Abakah-Dwumfour-MeftehWali-2022-ApplEcon}.

It is clear from the above literature that heavier and asymmetric tail dependence is prevalent in commodity futures, but the DY index is a conditional mean-based measure of the level of risk spillover.
Some scholars have already combined the traditional DY model with the quantile approach to study the risk spillover between commodity and financial markets. Based on this model, \cite{FJ-Chen-Liang-Ding-Liu-2022-EnergyEcon} find that spillovers between markets are stronger than those under average market conditions when considering extreme positive or negative events. \cite{FJ-Bouri-Lucey-Saeed-Vo-2020-IntRevFinancAnal} study the exchange rate market and obtain similar results, noting that tail risk premiums are asymmetric. However, the existing literature is mostly based on the traditional rolling-window DY risk spillover index rather than the more optimal TVP-VAR-DY model to combine with the quantile model.
Thus, our study considers risk spillover measurement based on TVP-VAR-DY with the inclusion of a quantile model to capture risk spillover in the upper and lower tails. 



\section{Methodology}
\label{S1:Methodology}

In recent years, the risk spillover index measure for networks of variables proposed by \cite{FJ-Diebold-Yilmaz-2009-EconJ,FJ-Diebold-Yilmaz-2012-IntJForecast} has been widely used. However, this method requires rolling window regression in dynamic applications, which has drawbacks such as the subjectivity brought by artificially setting the width of the window and the possible loss of sample information in the rolling window period. Based on the time-varying parameter vector autoregression (TVP-VAR) model, \cite{Antonakakis-Chatziantoniou-Gabauer-2020-RFM} improve and innovate the DY spillover index framework to obtain the TVP-VAR-DY model. 
Considering the extreme risk scenario, our study adds the quantile method to the TVP-VAR-DY, changing the VAR to a quantile VAR and using the method proposed by \cite{FJ-Koenker-Roger-1978-Econometrica} for the correlation estimation.

According to the Bayesian Information Criterion (BIC), the TVP-VAR(1) model can be written as follows, 
\begin{equation}
Y_t = \beta_{t}Y_{t-1}+\varepsilon_t, \qquad \varepsilon_t|\Omega_{t-1}\sim N(0,\Sigma_t)
\label{Eq:VAR}
\end{equation}
\begin{equation}
\beta_{t}=\beta_{t-1}+\xi_t, \qquad \xi_t|\Omega_{t-1}\sim N(0,\Xi_t)
\label{Eq:VAR parameters}
\end{equation}
\begin{equation}
Y_t = \sum_{j=0}^{\infty} {A_{jt}}{\Sigma_{t-j}}
\label{Eq:VMA}
\end{equation}
where $Y_t$, $Y_{t-1}$ and $\varepsilon_t$ are $N\times1$ dimensional vectors,  $\beta_{t}$,  $\xi_t$ and $\Sigma_t$ are $N\times{N}$ dimensional matrices, whereas $\Xi_t$ is $N^{2}\times{N^{2}}$ dimensional matrices, $\Omega_{t-1}$ represents all information available until ${t-1}$. 
After estimating the time-varying coefficients and variance-covariance matrices based on Kalman filter, we transform the TVP-VAR to its vector moving average (VMA) representation based on the Wold representation theorem in Eq.~(\ref{Eq:VMA}).

For the quantile-based TVP-VAR model, only the left-hand side of Eq.~(\ref{Eq:VAR}) needs to be changed,
\begin{equation}
  Q_{t,\tau}(Y_t|Y_{t-1},\cdots,Y_{t-p}) = \beta_{t}Y_{t-1}+\varepsilon_t,  \qquad \varepsilon_t|\Omega_{t-1}\sim N(0,\Sigma_t)
\label{Eq:quantile VAR}
\end{equation}
where $Q_{t,\tau}(Y_t|Y_{t-1},\cdots,Y_{t-p})$ is dependent variable of period $t$ at quantile $\tau(\tau\in(0,1))$. It is sufficient to re-estimate the coefficient matrix and the variance covariance matrix on this basis.

Our study estimates the risk spillovers based on the generalized impulse response function (GIRF) and generalized prediction error variance decomposition (GFEVD) as \cite{FJ-Diebold-Yilmaz-2014-JEconom}.
The first step is to compute the time-varying impulse response functions $GIRFs(\mit\Psi_{ij,t}(H))$ for all variables $j$ to variable $i$ following a shock.  The second step is to calculate the prediction error for the forward $H$ step when $i$ is shocked and unshocked, which is calculated as follows. 
\begin{equation}
GIRFt(\rm{H},\mit\delta_{j,t},\Omega_{t-1})=E(y_{t+H}|e_j=\mit\delta_{j,t},\Omega_{t-1})-E(y_{t+H}|\Omega_{t-1}),
\label{Eq:GIRF}
\end{equation}
\begin{equation}
\varPsi_{j,t}(H)={\frac {{A_{H,t}\Sigma_{t}}e_j} {\sqrt{\Sigma_{jj,t}}}}{\frac {\delta_{j,t}} {\sqrt{\Sigma_{jj,t}}}}, \qquad \delta_{j,t}=\sqrt{\Sigma_{jj,t}}, 
\label{Eq:psi1}
\end{equation}
and
\begin{equation}
\varPsi_{j,t}(H)=\Sigma_{jj,t}^{-1/2}A_{H,t}\Sigma_{t}e_j,
\label{Eq:psi2}
\end{equation}
where $\delta_{j,t}$ represents the selection vector, $\varPsi_{j,t}(H)$ represents the $GIRFs$ of variable $j$. The $H$-step ahead $GFEVD(\mit\phi_{ij,t}(H))$ can be calculated as 
follows, 
\begin{equation}
\bar{\mit\phi}_{ij,t}(H)=\frac {\sum_{t=1}^{H-1} {\varPsi_{ij,t}^2}} {\sum_{j=1}^{N}{\sum_{t=1}^{H-1} {\varPsi_{ij,t}^2}}}
\label{Eq:GFEVD}
\end{equation}
$\sum_{t=1}^{N}\bar{\mit\phi}_{ij,t}(H)=1$, $\sum_{i,j=1}^{N} \bar{\mit\phi}_{ij,t}(H)=N$. The GFEVD can be interpreted as the variance share one variable has on other variables $j$. 

Using the GFEVD, the total risk spillover index can be obtained,
\begin{equation}
Total_t(H)=C_t(H)=\frac {\sum_{i,j=1,i\ne{j}}^{N} {\bar{\mit\phi}_{ij,t}(H)}} {\sum_{i,j=1}^{N} {\bar{\mit\phi}_{ij,t}(H)}}\times 100=\frac {\sum_{i,j=1,i\ne{j}}^{N} {\bar{\mit\phi}_{ij,t}(H)}} {N}\times 100
\label{Eq:Total}
\end{equation}
The directional risk spillover index ($TO$) of $i$ over all other variables $j$ is calculated as follows: 
\begin{equation}
TO_{i,t}(H) = C_{i\to{\cdot},t}(H)=\frac {\sum_{j=1,i\ne{j}}^{N} {\bar{\mit\phi}_{ji,t}(H)}} {\sum_{j=1}^{N} {\bar{\mit\phi}_{ji,t}(H)}}\times 100
\label{Eq:TO}
\end{equation}
The directional risk spillover index ($FROM$) of all other variables $j$ to $i$ is calculated as follows: 
\begin{equation}
FROM_{i, t}(H)=C_{i\gets{\cdot},t}(H)=\frac {\sum_{j=1,i\ne{j}}^{N} {\bar{\mit\phi}_{ij,t}(H)}} {\sum_{i=1}^{N} {\bar{\mit\phi}_{ij,t}(H)}}\times 100
\label{Eq:FROM}
\end{equation}
The net risk spillover index ($NET$) of variable $i$ to $j$ is calculated as follows: 
\begin{equation}
NET_{i,t}(H)=TO_{i,t}(H)-FROM_{i,t}(H)
\label{Eq:fx11}
\end{equation}
If $C_{i,t}>0$, it means that variable $i$ influences network more than it is influenced by network, and is a net risk transmitter. If $C_{i,t}<0$, it means that variable i is driven by the network.

In addition, the measure of whether variable $i$ dominates other variables can be calculated in the following way: 
\begin{equation}
NPDC_{ij}(H)=\bar{\mit\phi}_{ji,t}(H)-\bar{\mit\phi}_{ij,t}(H)
\label{Eq:NPDC}
\end{equation}
\begin{equation}
DOM_{i,t}(H)=\sum_{j=1,i\ne{j}}^{N}\bm{\mathbf{I}}(NPDC_{ij,t}(H))
\label{Eq:DOM}
\end{equation} 
where $\bm{\mathbf{I}(\cdot)}$ is the indicator function, and $0\leq{DOM_{i,t}(H)}\leq{N-1}$.
If $NPDC_{ij}(H)>0$, it means that variable $i$ is risk spillover driven for variable $j$, and vice versa for variable $j$ is risk spillover driven for variable $i$. 
$DOM$ can be used to measure to how many other variables a given variable transmits risk.

\section{Data}
\label{S1:Data}

\subsection{Data source}

In order to study the risk spillover effects between different types of agricultural futures in the US and Chinese markets, we select the daily data of 11 staple food futures to study. The listing exchanges and countries of the chosen agricultural futures are presented as Table~\ref{Tab:dataset}. The price series is selected from July 9, 2014 to December 31, 2022, with data primarily from the Bloomberg database\footnote{https://www.bloomberg.com}, supplemented by the Wind database\footnote{https://www.wind.com.cn}. We use the daily return of agricultural futures to conduct empirical research. On the one hand, for different types of agricultural futures, price comparisons are not meaningful, and for agricultural futures in different markets, it is difficult to compare them because the units of valuation are not standardized. On the other hand, the rate of return is more important for investors as well as regulators, who need to be paid and who need to understand the upside and downside. In addition, agricultural futures yields have better statistical properties relative to prices. Considering the significant differences in the operation mechanism of different agricultural futures in different markets, for the missing values of data for agricultural futures, we use the closing price of the previous day as a supplement, and if the data of the previous day is still missing, the price of two days ago is filled in, and the rest is filled in the same way.

\begin{table}[!h]
  \renewcommand\arraystretch{1}
  \renewcommand\tabcolsep{22pt} 
   \centering
  \caption{Basic information of agricultural futures}
  \begin{tabular}{llccccc}
    \toprule
      Futures & Futures (abbr.)  & Exchange & Country & Listing date  \\ 
    \midrule
    Corn & Corn & CBOT   & USA & 2000/01/01  \\
    Rough Rice & RiceR & CBOT  & USA & 2003/11/19 \\
    Soybean & Soybean & CBOT  & USA & 2000/01/04 \\
    Wheat & Wheat & CBOT  & USA & 2000/01/01 \\
    Wheat WH & WheatWH & CZCE & China & 2003/03/28 \\
    Early Rice & RiceE & CZCE & China & 2009/04/20 \\
    Wheat PM & WheatPM & CZCE & China & 2012/01/17 \\
    Japonica Rice & RiceJ & CZCE & China & 2013/11/18 \\
    Late Indica Rice  & RiceLI  & CZCE & China & 2014/07/08 \\
    No. 1 Soybean  & Soybean  & DCE & China& 2000/01/04 \\
    Corn & Corn & DCE & China & 2004/09/22 \\
    \bottomrule
  \end{tabular}
  \label{Tab:dataset}
\end{table}

\subsection{Data description}

To obtain the risk spillovers of agricultural futures, the daily logarithmic return on $\Delta{t}$ is first calculated for the $i$th agricultural futures:
\begin{equation}
   r_{i}(t)=\ln(P_{i}(t))-\ln(P_{i}(t-\Delta{t}))
   \label{Eq:fx14}
\end{equation}
where $r_{i}(t)$ and $P_{i}(t)$ are respectively the daily return and daily closing price of the $i=1,\,\cdots,\, N$ agricultural futures on day $t=1,\,\cdots,\, T$.

Table~\ref{Tab:description} reports the summary statistics of agricultural futures. Apart from rice agricultural futures, the average values of all types of agricultural futures are positive, and the standard deviation ranges from $[0.004, 0.008]$. As can be seen in Table~\ref{Tab:description}, all agricultural futures returns are either left-biased or right-biased, with a higher percentage of left-biased, and the kurtosis of agricultural futures returns is greater than 3 for all agricultural futures except CBOT wheat, suggesting that the overall empirical return segments show a phenomenon of fat tail. The Jarque-Bera test shows that the returns do not follow a normal distribution.

\begin{table}[!h]
  \renewcommand\arraystretch{1}
  \renewcommand\tabcolsep{6.1pt} 
   \centering
  \caption{Descriptive statistics of agricultural futures }
  \begin{tabular}{llllllll}
    \toprule
        & Mean($\times 10^4$) & SD & Skewness & Kurtosis & J-B & ADF \\ 
    \midrule
    CBOT Corn & $1.137424$  & $0.006596$ & $-1.611125$  & $22.669498$ & $47277 ^{***}$& $-34.7506^{***}$  \\
    CBOT Rough Rice & $-0.033893$ & $0.006968$ &$-10.713240$  & $313.655834$ & $9072653^{***}$ & $-18.1016^{***}$  \\
    CBOT Soybean & $0.820774$ & $0.005672$ & $-1.322053$  & 12.818655 & $15721^{***}$ & $-8.8552^{***}$ \\
    CBOT Wheat & $0.669145$ & $0.007941$ & $0.287490$  & $2.195061$ & $472^{***}$ & $-47.6971^{***}$ \\
    CZCE Wheat WH & $0.490771$ & $0.006912$ & $-0.804597$  & $45.066920$ & $186649^{***}$ & $-10.6576^{***}$ \\
    CZCE Early Rice &$-0.128320$ & $0.004768$ & $-3.203338$  & $64.527406$ & $385951^{***}$ & $-21.0501^{***}$ \\
    CZCE Wheat PM & $0.255962$ & $0.005511$ & $1.150434$  & $27.937186$ & $72116^{***}$ & $-12.6489^{***}$ \\
    CZCE Japonica Rice & $-0.408400$ & $0.005993$ & $1.558288$ & $29.814074$ & $82472^{***}$ & $-12.4500^{***}$ \\
    CZCE Late Indica Rice & $-0.561970$ & $0.004716$ & $-0.402076$  & $38.483367$ & $135982^{***}$ & $-15.7857 ^{***}$ \\
    DCE No.1 Soybean & $0.079897$ & $0.004400$ & $0.614403$  & $10.256078$ & $9789^{***}$ & $-44.9036^{***}$ \\
    DCE Corn & $0.012269$ & $0.004547$ & $-2.897030$ & $69.676683$ & $448694^{***}$ & $-7.9324^{***}$ \\
    \bottomrule
  \end{tabular}
  \begin{tablenotes}    
        \footnotesize        
        \item[1] Note: J-B is the Jarque-Bera test, and ADF is the ADF test. $^{***}$ denotes statistical significance at the 1\% confidence level.
      \end{tablenotes}            
  \label{Tab:description}
\end{table}

In addition, this study applies ADF to test the stationarity of the time series. The ADF test show that the return rejects the hypothesis of the existence of a unit root at the 1\% significance level, and it is a stationary time series, which can be used for the next step in the construction of the spillover index. Subsequently, the optimal lag order of the model is determined to be the 1st order according to the BIC. Therefore, our study establishes the 11-variable lagged 1st order TVP-VAR model with the quantile model.

\section{Empirical analysis}
\label{S1:EmpAnal}

\subsection{Conditional mean risk spillover effects}

The empirical results of the static risk spillover effects of the study are presented in the form of a table of $11\times{11}$. The non-diagonal elements of the matrix represent the risk spillover of the variable pointed to by the column to the variable pointed to by the row, $TO$ and $FROM$ are the directional spillover indices, and the $NET$ row is the net spillover index of this agricultural commodity futures, which is obtained by subtracting $FROM$ from $TO $ of the variable. The lower-right $TCI$ is the static total spillover index, which represents the average risk premium across all variables.

Table~\ref{Tab:directional spillovers} shows the static risk spillovers among agricultural futures. Overall, the average risk spillover level $TCI$ across all variables is 18.8\%. The $TO$ index has a wider range than the $FROM$ index, with the results of the $TO$ index ranging from 8\% to 50.9\% and the $FROM$ index ranging from 6.7\% to 42\%. Specifically, according to the results of the $TO$ index, the three agricultural futures with the largest risk transmitting are CBOT Corn, CBOT Soybean, and CBOT Wheat, and the three agricultural futures with the smallest risk transmitting are CZCE Late Indica rice, CZCE Wheat PM, and CZCE Japonica rice, which are all less than 10\%. 
According to the results of the $FROM$ index, the three agricultural futures with the highest risk receiving are still CBOT corn, CBOT soybean, and CBOT wheat, but the level is not as high as that of the $TO$ index, so the net spillover risk level of the three agricultural futures mentioned above is very high, and the ones with the lowest risk receiving are CZCE Late Indica Rice, CZCE Wheat WH, and CZCE Japonica Rice. 
Taken together, the above results indicate that CBOT Corn, CBOT Soybeans, and CBOT Wheat play a significant role as the primary risk transmitter and receiver in the risk system between China and the US agricultural futures, while CZCE Japonica Rice is at the edge of the agricultural futures risk system, with a lower level of both transmitted and received, which is related to its trading volume and liquidity. 

The net spillover index $NET$ can show how each agricultural futures contract contributes to risk spillover. CBOT corn is the largest risk transmitter, with a $NET$ index of 8.8\%, followed by CBOT soybean and CBOT wheat, also significant risk transmitters, with a $NET$ index of 5.2\% and 2.4\%, respectively. The top two negative $NET$ index agricultural futures are DCE soybean and DCE corn, with $-$6.8\% and $-$5.4\%, respectively, indicating that they are risk receivers. In addition, with the exception of the five agricultural futures with the highest levels of the aforementioned $NET$ index, all of the remaining agricultural futures' net spillover indices are within 3\% of one another, suggesting that the majority of them are at an equilibrium level of risk connectedness.

The net risk transfer number $DOM$ can indicate how many other agricultural futures receive the risk transmitted by the given agricultural futures, and thus $DOM$ can be applied to support the risk role of the agricultural futures. In Table \ref{Tab:directional spillovers}, CBOT Corn has a net risk spillover to 9 agricultural futures, CBOT Wheat and Soybean have a net risk spillover to 8 and 7 agricultural futures, respectively, while CZCE Early Indica Rice, DCE No.1 Soybean, and Corn have a net risk spillover to just one agricultural commodity.

From the perspective of agricultural product types, those that play a major role in risk spillover are corn, soybean, and wheat agricultural product futures, and rice agricultural product futures are lower in both directional risk spillover level and net risk spillover level, suggesting that rice does not have a sharp response to external risk shocks.This may be related to the fact that regions where rice is the main grain will adopt active grain price stabilization policies, such as the minimum purchase price policy for grain implemented in China.  Based on this characteristic of rice, it can be considered a better risk-aversion option.

\begin{table}[thp]
  \renewcommand\arraystretch{1}
  \renewcommand\tabcolsep{2pt} 
   \centering
  \caption{Static risk spillover under conditional mean}
  \begin{tabular}{lccccccccccccccc}
    \toprule
        & \multicolumn{4}{c}{CBOT} &   \multicolumn{5}{c}{CZCE} &   \multicolumn{2}{c}{DCE} & \multirow{2}{*}{FROM} \\
        \cmidrule(lr){2-5}  \cmidrule(lr){6-10} \cmidrule(lr){11-12}
        & Corn & RiceR  & Soybean &  Wheat & WheatWH & RiceE  & WheatPM & RiceJ  & 
        RiceLI  & Soybean & Corn & \\ 
    \midrule
    
   CBOT Corn & 58.0 & 1.7 & 18.1 & 18.3 & 0.7 & 0.4 & 0.9 & 0.5 & 0.4 & 0.5 & 0.5 & 42 \\ 
   CBOT RiceR & 2.3 & 87.6 & 2.4 & 2.0 & 0.8 & 0.6 & 0.7 & 0.8 & 0.7 & 0.9 & 1.3 & 12.4 \\ 
   CBOT Soybean & 19.8 & 1.8 & 64.0 & 9.1 & 1.0 & 0.7 & 0.9 & 0.9 & 0.4 & 1.0 & 0.5 & 36 \\ 
   CBOT Wheat & 20.3 & 1.6 & 9.4 & 64.3 & 0.8 & 0.7 & 0.5 & 0.7 & 0.4 & 0.5 & 0.7 & 35.7 \\ 
   CZCE WheatWH & 0.9 & 0.6 & 0.9 & 1.1 & 90.4 & 1.4 & 0.7 & 0.8 & 1.3 & 1.1 & 0.9 & 9.6 \\ 
   CZCE RiceE & 0.9 & 0.7 & 1.4 & 1.0 & 1.6 & 87.7 & 1.4 & 1.6 & 1.7 & 1.1 & 0.7 & 12.3 \\ 
   CZCE WheatPM & 1.3 & 1.1 & 1.2 & 0.8 & 0.7 & 1.5 & 89.8 & 0.9 & 1.1 & 0.8 & 0.7 & 10.2 \\ 
   CZCE RiceJ & 0.9 & 0.7 & 1.1 & 1.0 & 1.1 & 1.4 & 1.0 & 90.3 & 0.6 & 1.3 & 0.6 & 9.7 \\ 
   CZCE RiceLI & 0.4 & 0.6 & 0.3 & 0.6 & 1.0 & 1.2 & 0.8 & 0.5 & 93.3 & 0.6 & 0.6 & 6.7 \\ 
   DCE Soybean & 1.8 & 1.1 & 4.3 & 1.9 & 1.4 & 1.1 & 1.0 & 1.2 & 0.8 & 82.5 & 2.9 & 17.5 \\ 
   DCE Corn & 2.3 & 1.3 & 2.1 & 2.1 & 1.0 & 0.7 & 0.7 & 1.1 & 0.6 & 2.8 & 85.4 & 14.6 \\ 
   TO & 50.9 & 11.2 & 41.3 & 38.1 & 10.1 & 9.9 & 8.4 & 8.9 & 8.0 & 10.6 & 9.3 &  \\ 
   NET & 8.8 & $-$1.2 & 5.2 & 2.4 & 0.6 & $-$2.4 & $-$1.8 & $-$0.8 & 1.3 & $-$6.8 & $-$5.4 & TCI \\ 
   DOM & 9.0 & 3.0 & 7.0 & 8.0 & 6.0 & 1.0 & 2.0 & 3.0 & 7.0 & 1.0 & 1.0 & 18.8 \\ 
    \bottomrule
  \end{tabular}
  \begin{tablenotes}   
        \footnotesize          
        \item[1] Note: This study conducts the generalized forecast error variance decomposition with forecast steps being 5 days. The same is true below.
      \end{tablenotes}           
  \label{Tab:directional spillovers}
\end{table}

\subsection{Conditional mean dynamic risk spillover effects}

\subsubsection{Dynamic total risk spillovers}

\begin{figure}[h!]
    \centering
    \includegraphics[width=0.6\linewidth]{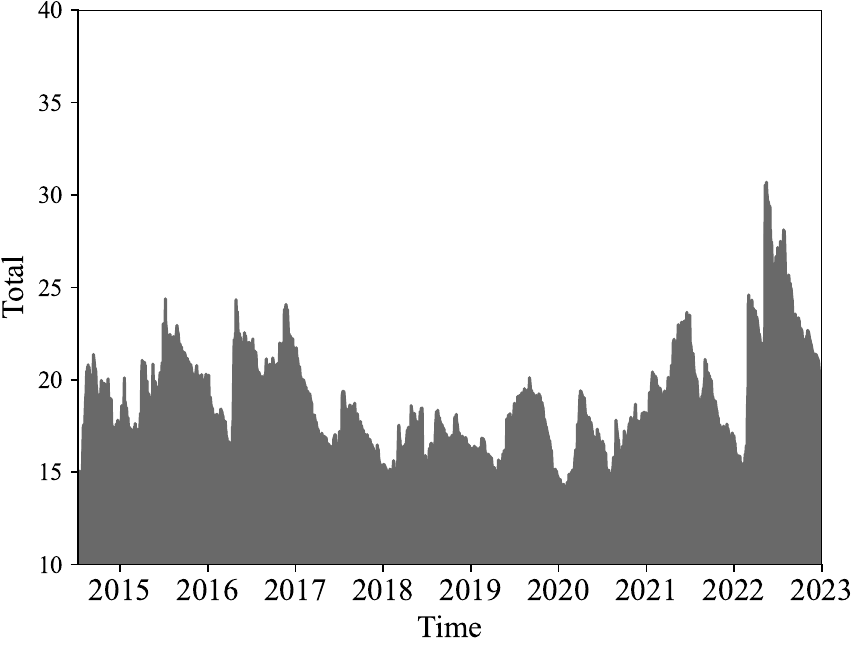}
    \vspace{-4mm}
    \caption{Total spillover under conditional mean}
    \label{Fig:AgroFutures_TVP_VAR_DY_TOTAL}
\end{figure}

Fig.~\ref{Fig:AgroFutures_TVP_VAR_DY_TOTAL} illustrates the dynamic total risk spillovers among agricultural futures at the conditional mean. The dynamic total risk spillovers index enables us to comprehend how total risk spillovers have changed over time and to identify the impact of shocks from various political or economic events.

As shown in Fig.~\ref{Fig:AgroFutures_TVP_VAR_DY_TOTAL}, the total risk spillovers among agricultural futures markets are significant and time-varying, with a clear upward trend in volatility after 2018, which is related to the globalization of the economy and the rise in political and economic instability. The volatility of the total risk spillover index ranges from 10\% to 30\%.

It is evident from Fig.~\ref{Fig:AgroFutures_TVP_VAR_DY_TOTAL} that significant total risk spillover occurred in 2015, 2021, and 2022. 
In 2015, global daily crude oil production increased sharply, and the average settlement prices of Brent and WTI futures fell 47\% and 48\% each year-on-year. 
There is a significant dependence pattern of information spillover between crude oil and agricultural markets \citep{FJ-Hung-2021-ResourPolicy}, and the net volatility spillover of crude oil to agricultural products tends to increase when crude oil remains in a relatively high state of volatility \citep{FJ-Yip-Brooks-Do-Nguyen-2020-IntRevFinancAnal}. Along with China's economic downturn and stock market crash, agricultural futures were strongly affected.
In 2021, due to the COVID-19 pandemic, food production in many countries is reduced and accompanied by related agricultural export restriction policies in food-exporting countries, resulting in food panic and an increase in the total risk spillover.
In 2022, the Russian-Ukrainian conflict broke out, while Russia and Ukraine are the world's major food producers and exporters. In 2021, the two countries exported a total of 62.85 million tons of wheat, accounting for 32.5\% of the global wheat exports. The Russia-Ukraine conflict affects the supply of the remaining agricultural importing countries, and risk transmission causes a substantial rise in the total risk spillover index.

\subsubsection{Dynamic directional risk spillovers}

The total risk spillover index can only reveal the overall level of risk spillover in the agricultural futures market, but it cannot capture the change in risk spillover of subdivided agricultural futures in the time domain, which is not conducive to the observation of the net spillover risk of subdivided agricultural futures, and thus requires the dynamic directional risk spillover index.
Dynamic directional risk spillover index, which can be divided into $TO$ and $FROM$, is a continuous measure of static directional spillover index at various time points and is primarily used to examine the change of spillover and symmetry of different agricultural futures.

\begin{figure}[!h]
    \centering
\includegraphics[width=0.975\linewidth]{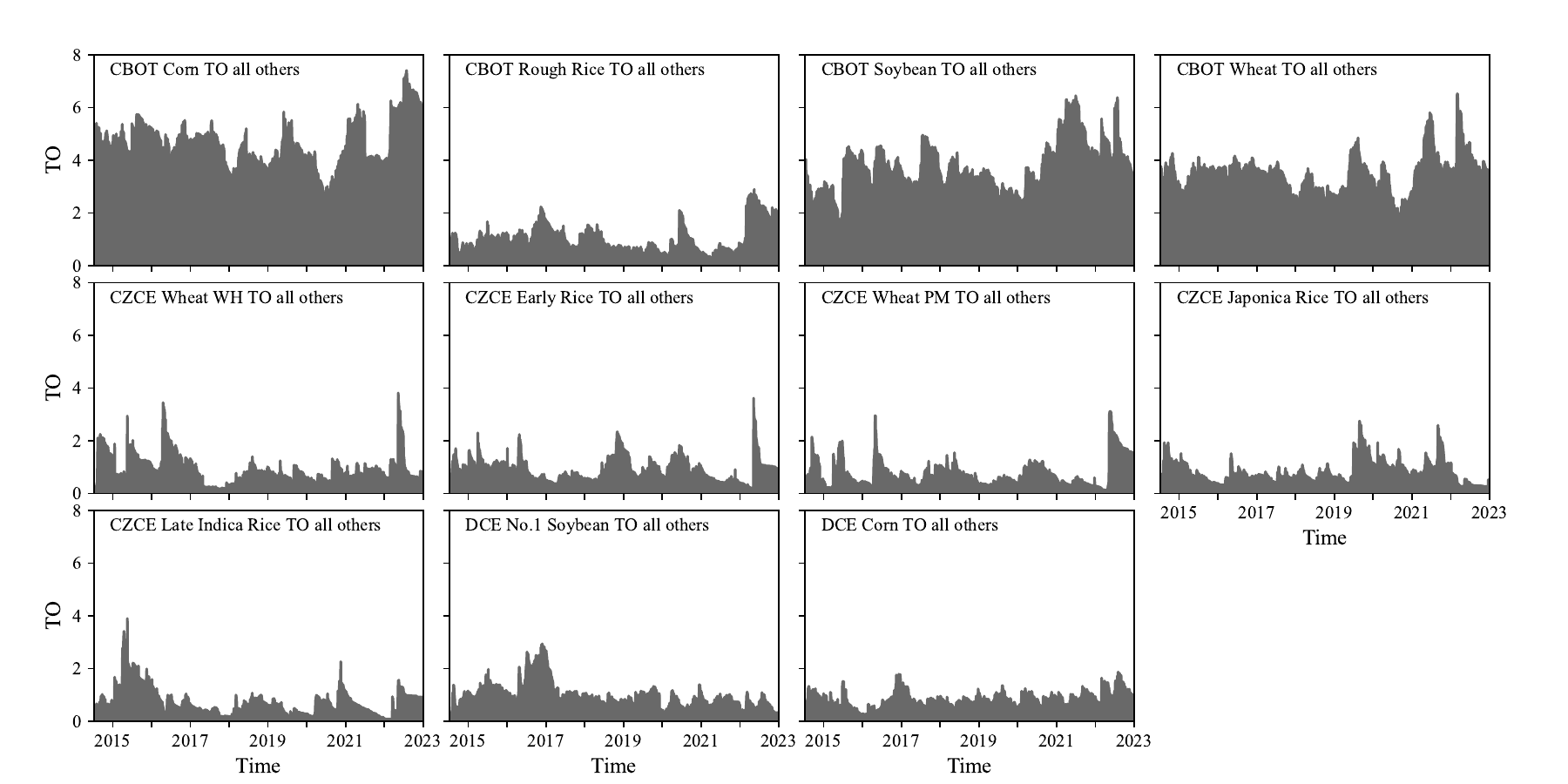}
    \caption{$TO$ spillover under conditional mean}
    \label{Fig:AgroFutures_TVP_VAR_DY_TO}
\end{figure}

\begin{figure}[!h]
    \centering
    \includegraphics[width=0.975\linewidth]{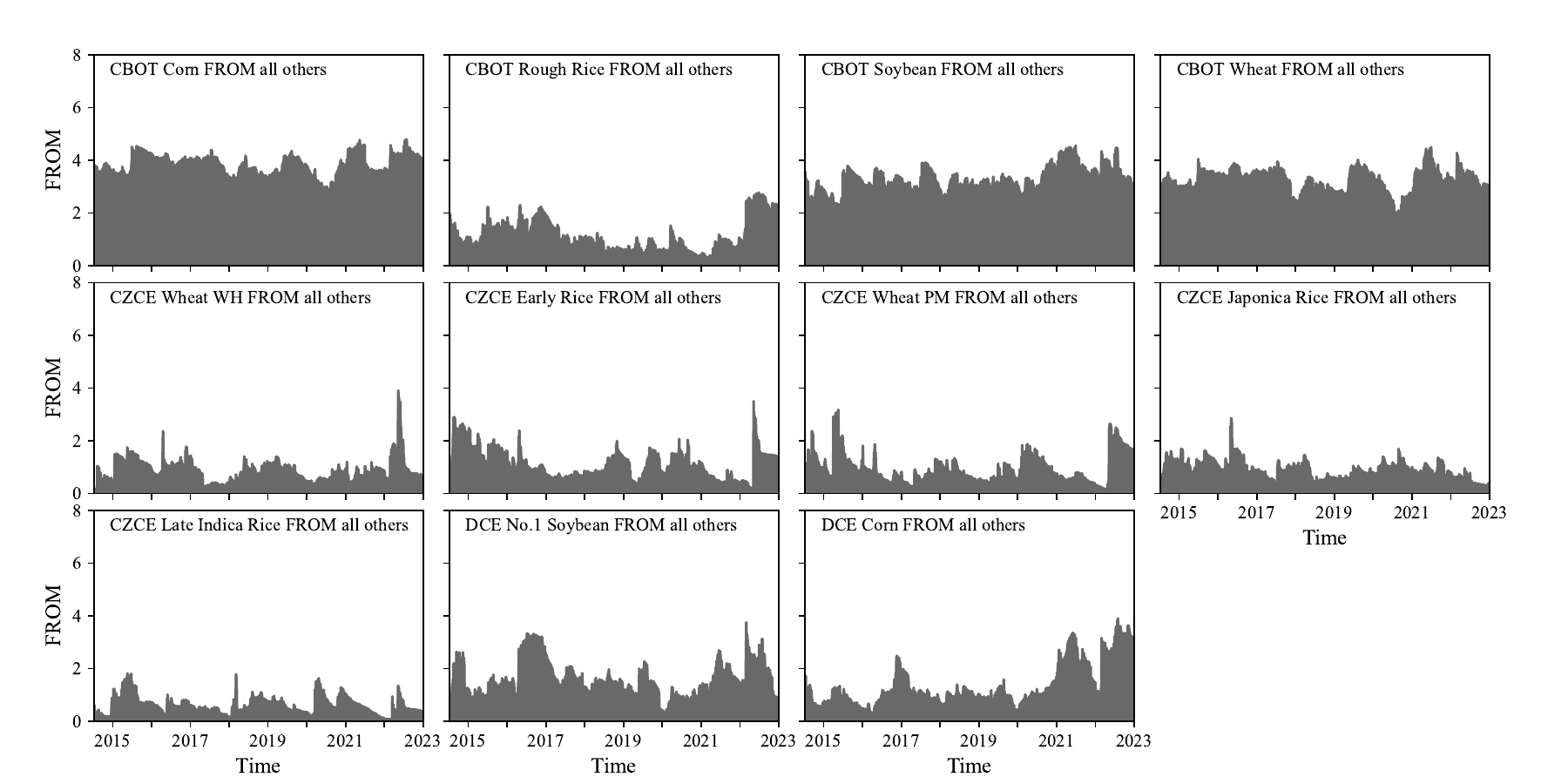}
    \caption{$FROM$ spillover under conditional mean}
    \label{Fig:AgroFutures_TVP_VAR_DY_FROM}
\end{figure}

Fig.~\ref{Fig:AgroFutures_TVP_VAR_DY_TO} and Fig.~\ref{Fig:AgroFutures_TVP_VAR_DY_FROM} show the dynamics of $TO$ and $FROM$ for 11 agricultural futures, respectively.
For both $TO$ and $FROM$, CBOT Corn, CBOT Soybean, and CBOT Wheat have the highest levels of risk spillovers. The $FROM$ index of DCE No. 1 soybean and DCE Corn is also high, far exceeding the $TO$ index of both, which is consistent with the conclusion that the net spillovers of DCE No. 1 soybean and DCE Corn are negative and large in Table~\ref{Tab:directional spillovers}.

From a time-domain perspective, Fig.~\ref{Fig:AgroFutures_TVP_VAR_DY_TO}and Fig.~\ref{Fig:AgroFutures_TVP_VAR_DY_FROM}show that the directional risk spillovers in agricultural futures markets are time-varying and volatile. The overall trend shows that the risk spillovers of agricultural futures are comparable. The $FROM$ index growth of DCE Corn has a more distinct trend from 2020, which is due to the surge in China's corn imports caused by the corresponding policy on corn as well as the demand for corn. For China, rice and wheat are food grains, and corn is a feed grain, which is in a relatively weaker position in terms of its importance. Additionally, China's corn planting was constrained in 2020 as an outcome of the COVID-19 epidemic, while hog farming had resumed and feed consumption had increased. As a result, corn imports skyrocketed, growing at a pace of 135.91\% year on year. 

Analyzing the significant volatility of different agricultural futures, it can be concluded that the 2020 COVID-19 pandemic and the 2022 Russia-Ukraine conflict have had a certain impact on the risk spillover of each agricultural futures.The COVID-19 pandemic increases the level of risk spillovers mainly through panic and export restrictions in some countries. The increase in risk spillovers from the Russian-Ukrainian conflict is mainly caused by the reduction in food production resulting from the war between the two countries, which affects the volume and structure of global supply.
For CBOT Soybean, the directional spillover in 2020 began to have a more substantial increase. Excluding the epidemic factor, the US trade war between China and the United States reached a first-phase agreement to promote the export of US soybeans to the world's largest soybean consumer, China, which has also increased the level of its risk spillover to a certain extent.
Compared with CBOT Soybean, CBOT Corn,and CBOT Wheat's risk spillover level in 2019-2020 years has a more obvious "V" trend, that is, in 2019, two agricultural futures have a larger risk spillover, mainly due to the US climate impact, resulting in a general delay in the corn sowing, a low wheat planting area, and superimposed on the Russian Federation and other countries to capture the global market share of wheat, and other factors, causing a high level of risk.

Compared with the 2020 COVID-19 pandemic, the impact of the 2022 Russia-Ukraine conflict will have a wider reach. As Russia is the world's third-largest producer and top exporter of wheat, and Ukraine is the world's third-largest exporter of wheat and the world's fourth-largest exporter of corn, the conflict brings about a significant increase in the level of risk spillovers in agricultural futures for corn and wheat. Following the outbreak of the conflict, Corn and wheat agricultural futures have all seen steeper increases in spillover levels. In addition, CBOT Rice and CZCE Early Indica Rice also have a significant increase in the level of spillover premium, mainly due to two reasons. First, the frequent occurrence of global climate extremes, the drought in the southern hemisphere, the extreme cold and flooding in the northern hemisphere, planting of the unfavorable environment brought about by the decline in yields, and the second reason is that the price volatility of other agricultural commodities has a certain degree of risk spillover on rice. 

\subsubsection{Net risk spillovers}

\begin{figure}[!h]
    \centering
    \includegraphics[width=0.975\linewidth]{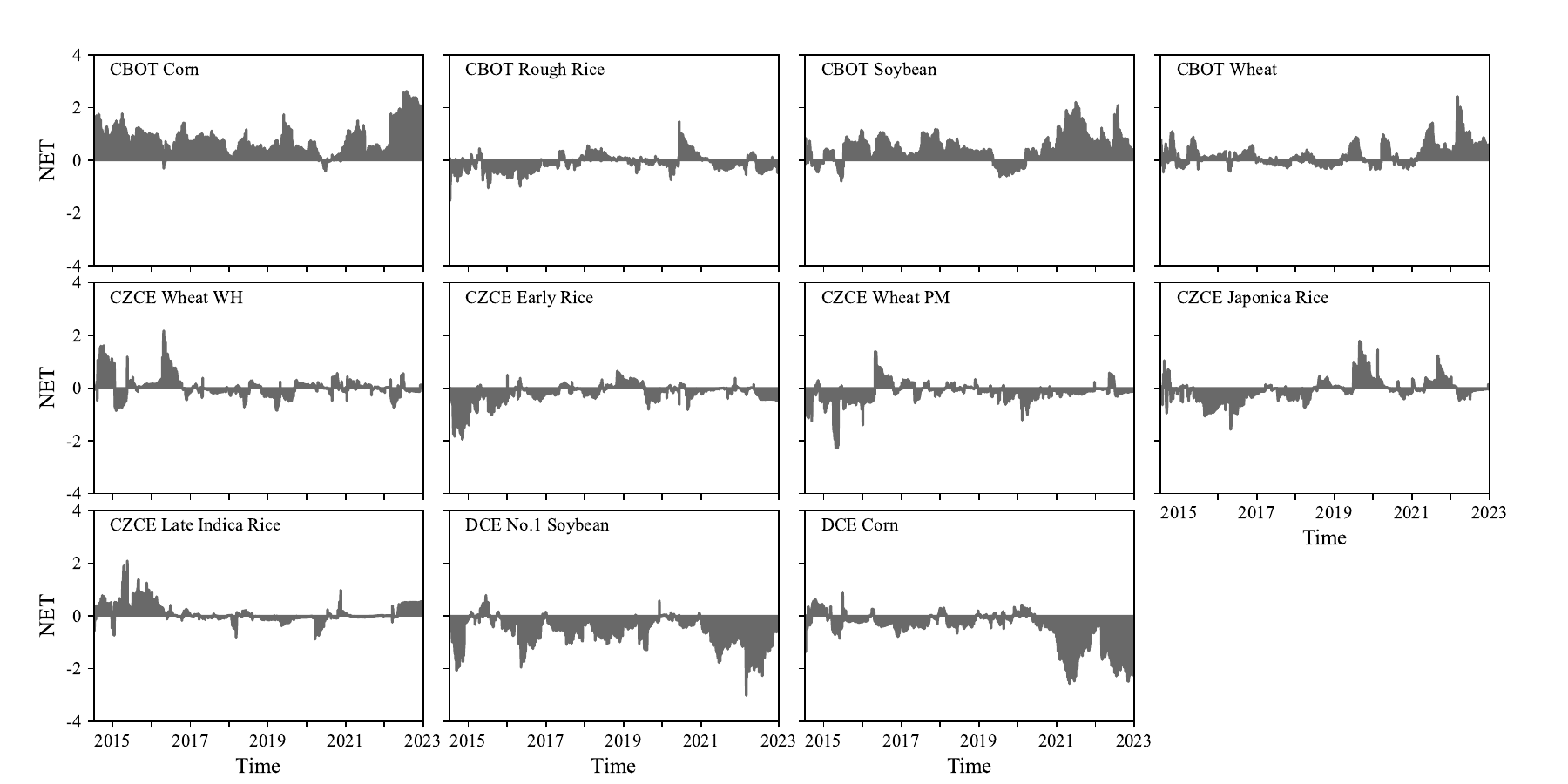}
    \caption{$NET$ spillover under conditional mean}
    \label{Fig:AgroFutures_TVP_VAR_DY_NET}
\end{figure}

Net risk spillover combines the $TO$ index and $FROM$ index of each agricultural commodity futures to obtain the different roles of each agricultural commodity futures in the risk spillover process. Fig.~\ref{Fig:AgroFutures_TVP_VAR_DY_NET} presents the dynamic net spillovers for agricultural futures at the conditional mean.
According to Fig.~\ref{Fig:AgroFutures_TVP_VAR_DY_NET}, CBOT Corn, CBOT Soybean, and CBOT Wheat play a significant role in risk spillovers as net risk transmitters, with CBOT Corn being the most important risk transmitter. DCE No.1 Soybean and DCE Corn are the main net risk receivers, with DCE Corn's net risk receiving tendency being prominent from 2020.
For obvious net risk transmitters and net risk receivers across the sample period, the net risk spillover index tends to rise in absolute terms, but for other agricultural futures where the roles are less evident, the net risk premium remains modest. In the context of heightened economic uncertainty and agricultural risk, it implies that there is some centrality and aggregation of spillovers, and that risk is primarily contagious among many key agricultural futures.

A comparison of risk spillovers across different agricultural futures categories reveals that, on average, corn and soybeans play a major role in agricultural futures risk premiums, followed by wheat, and finally rice. This is mostly due to the fact that governments would prioritize the production of wheat and rice and regulate the volume and proportion of imports in order to ensure food security. In contrast to wheat and rice, corn and soybeans are generally considered to be feed grains.  At the same time, corn and soybeans can be used to produce biodiesel, which can be seen as an alternative to crude oil. As a result, external economic shocks can cause risk contagion to corn and soybeans first, causing greater price volatility, and leading to further risk contagion across agricultural futures.

Comparing the risk spillover of agricultural futures listed on different exchanges, it can be found that the transmitters of agricultural futures in the risk spillover are all CBOT-listed futures, while the receivers of risk are all DCE-listed futures. Expanding to the listed countries, the conclusion suggests that on average, US agricultural products have a strong position in risk spillovers, while China's agricultural products, especially soybeans and corn, are in a passive position. This is mainly due to the fact that the US is a major exporter of soybeans and corn, while China is a major importer of soybeans and corn.

\subsection{Conditional mean risk spillover network and minimum spanning tree }

\subsubsection{Risk spillover network}

\begin{figure}[h!]
    \centering
    \includegraphics[width=0.473\linewidth]{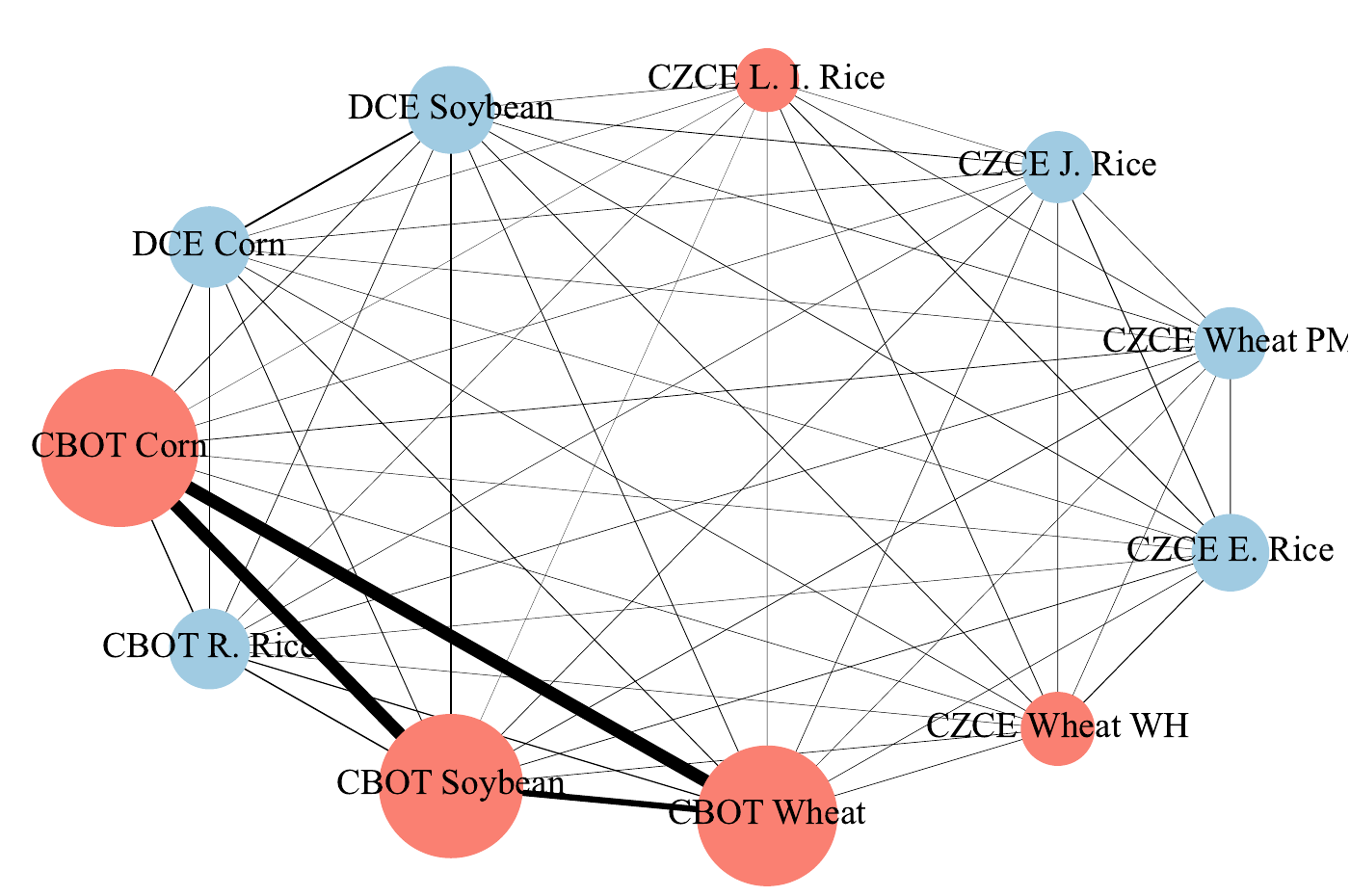}
    \includegraphics[width=0.473\linewidth]{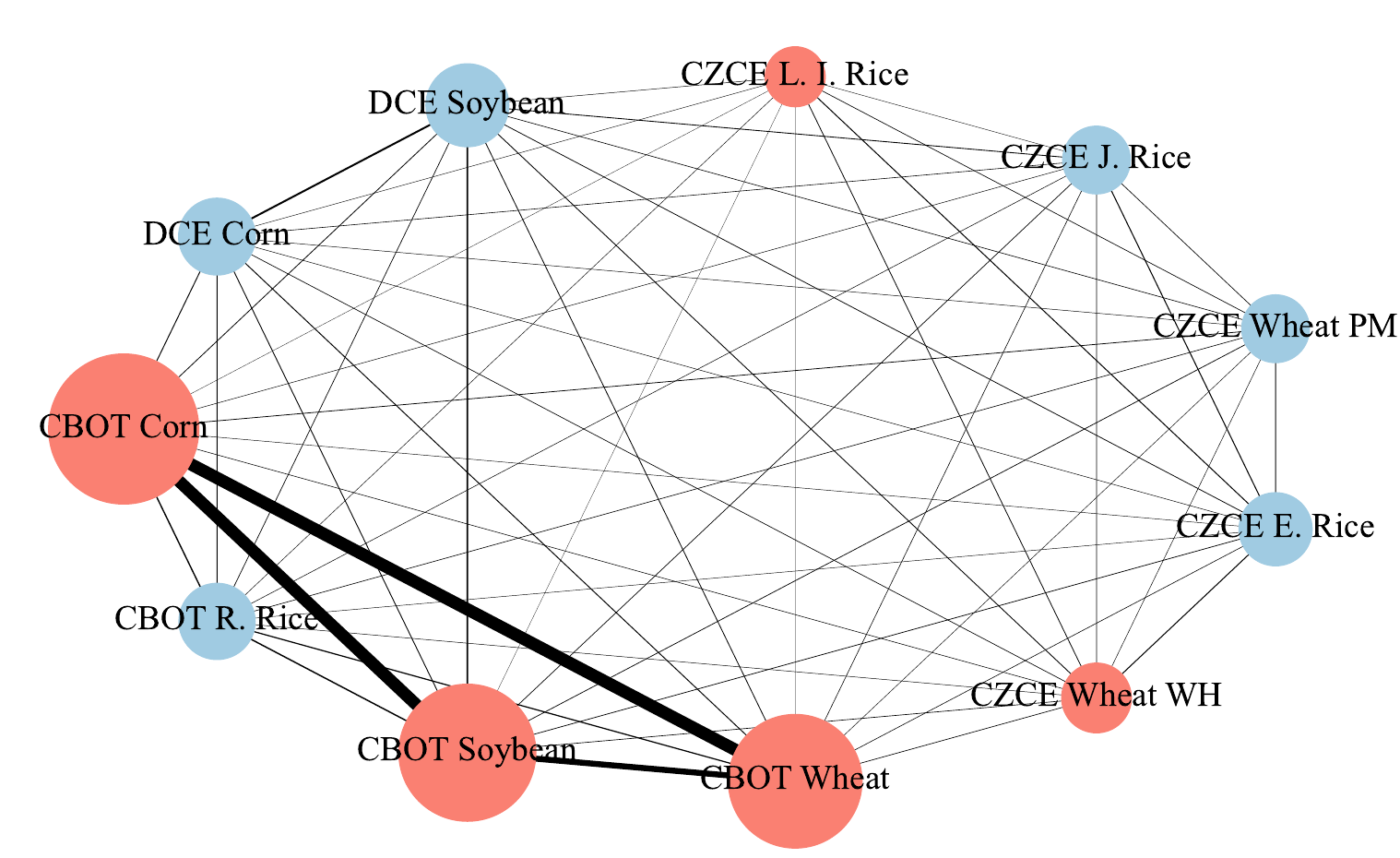}
    \caption{Spillover correlation network and net risk spillover network under conditional mean}
    \label{Fig:AgroFutures_TVP_VAR_DY_network}
\end{figure}

The correlation network is a network established by taking the sum of pairwise risk spillover among agricultural futures as the edge weights, which is a weighted undirected network, and the size of the node is determined by its average out-strength. The larger the node, the more strongly correlated the agricultural futures represented by it are with other nodes in risk spillover, and the thickness of the link between the nodes represents the strength of the correlation between the two nodes. The net spillover network is a weighted directed network built by using the pairwise net spillover index between agricultural futures as link weights. Our study considers the observation convenience and only retains the edges with net spillover values greater than 0.5\%. The size of the nodes is determined by their out-strength: the larger the node, the greater the risk spillover effect on other nodes, and the thickness of the links represents the magnitude of the value of the net spillover risk effect. In addition, red nodes represent positive net spillover risk effects, i.e., risk transmitters, and blue nodes represent negative net spillover risk effects, i.e., risk receivers.
Fig.~\ref{Fig:AgroFutures_TVP_VAR_DY_network} displays both the correlation and net risk spillover networks at the conditional mean.

The correlation network can be used to understand the strength of the relationship between each agricultural futures in the risk spillover. According to the left panel of Fig.~\ref{Fig:AgroFutures_TVP_VAR_DY_network}, we can find that there is a strong link between CBOT Corn, CBOT Wheat, and CBOT Soybean, and the correlation between the three is mainly dependent on CBOT Corn. The strong relationship between CBOT corn and CBOT soybeans is mainly formed by the substitutability of corn and soybeans, and the correlation between CBOT corn and CBOT wheat is probably formed because corn and wheat are the two most traded agricultural futures, resulting in market expectations  affecting both agricultural commodities at the same time. The link between CBOT Wheat and CBOT Soybean is relatively weak, but it still far exceeds the correlation between the rest of the agricultural futures.

The right panel of Figure \ref{Fig:AgroFutures_TVP_VAR_DY_network} shows the net risk spillover network. From the figure, we can see that the main risk transmitters are CBOT Corn, CBOT Wheat, and CBOT Soybean, and the risk between the three is transmitted from CBOT Corn to the remaining two agricultural futures. Although the level of risk spillover between CBOT Soybean and CBOT Wheat is high, the net spillover is processed to result in offsetting each other, suggesting that the two are in about the same position in risk spillover. CBOT Corn, Soybean, and Wheat all have strong risk spillovers to DCE Corn and DCE No.1 Soybean, with CBOT Soybean having the largest risk spillover to DCE No.1 Soybean. In addition, the network more clearly demonstrates that rice futures have some risk segregation from risk spillover and are, on average, the agricultural futures that suffer the least from risk spillover.

\subsubsection{Minimum spanning tree}

Our study employs the minimum spanning tree method to examine the risk spillover paths. We therefore take into consideration utilizing the matrix of the inverse of the net risk spillover matrix to design the minimum spanning tree because the bigger the net risk spillover index, the more significant the agricultural futures are in the risk spillover. Only one spillover link can be kept between the two agricultural futures when connecting connections are built using the minimum spanning tree method; this spillover path thus reflects the most likely spillover path.

\begin{figure}[!h]
    \centering
    \includegraphics[width=0.50\linewidth]{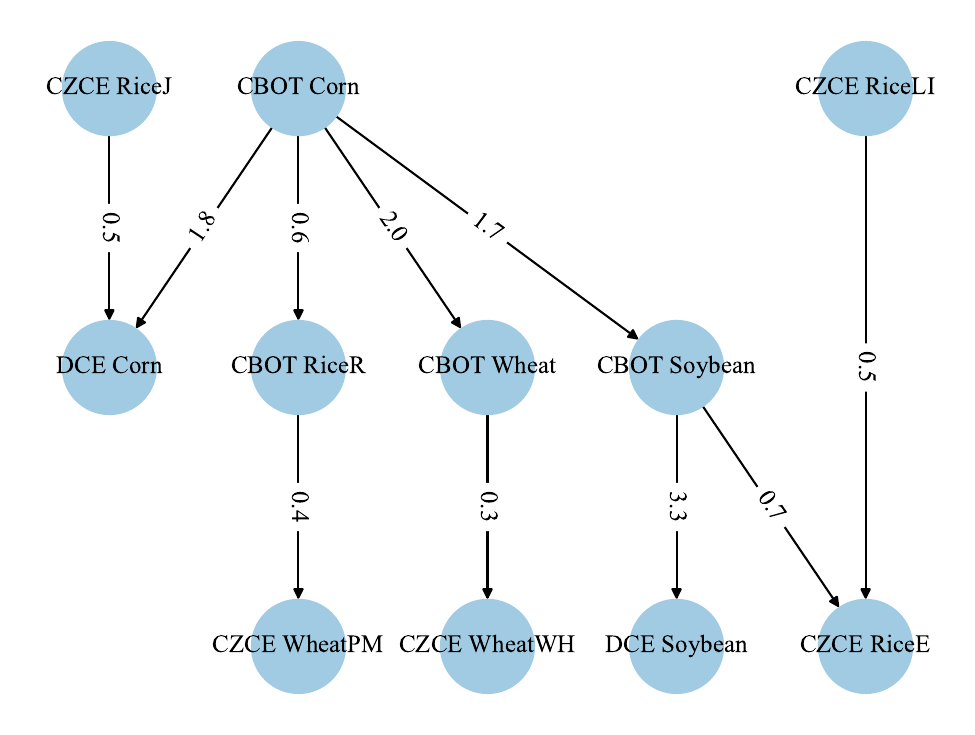}
    \caption{Minimum spanning tree under conditional mean}
    \label{Fig:AgroFutures_TVP_VAR_DY_MST}
\end{figure}

Fig.~\ref{Fig:AgroFutures_TVP_VAR_DY_MST} shows the minimum spanning tree graph for the sample period, with the node arrows representing the direction of risk spillover, and the numbers on the connections representing the magnitude of the pairwise net spillover risk between the two agricultural futures. From the minimum spanning tree diagram, it can be seen that the risk spillover among agricultural futures can be divided into three main tiers, with the first tier dominated by CBOT Corn, the second tier dominated by the rest of the CBOT-listed agricultural futures as well as DCE Corn, and the third tier dominated by the other agricultural futures. The risk transmission paths are all relatively short, and the longest transmission path includes only three nodes, indicating that the risk spillover among agricultural futures presents a clear center-dominated feature. Through the analysis of the minimum spanning tree, we find that CBOT Corn is the most important agricultural futures in the risk spillover, and CBOT Soybean has the tightest risk spillover to DCE 
No.1 Soybean, and US-listed agricultural futures are generally located at the top of the path, whereas China-listed agricultural futures are generally located at the end of the path of the minimum-spanning tree.

\subsection{Quantile risk spillover effects}

The presentation of the static risk spillover under quantiles is consistent with the presentation of the static risk spillover results under conditional averaging, which are tabulated. The tables have the same meaning. Table~\ref{Tab:directional spillovers quantile 0.5} shows the conditional median ($\tau=0.5$) static risk spillovers obtained from the quantile-based model. The conditional median static risk spillovers results are highly similar to the conditional mean static risk spillovers presented in Table~\ref{Tab:directional spillovers}, but the overall risk spillovers level $TCI$ reaches 21.9$ \%$, which is slightly higher than the 18.8$ \%$ under the conditional mean.

In order to distinguish the price risk spillovers between the average and extreme cases, our study estimates quantile risk spillovers for the extreme lower and upper quantiles. The spillover effects at the 5 percent ($tau=0.05$) and 95 percent ($tau=0.95$) quantiles are shown in Table~\ref{Tab:directional spillovers quantile 0.05} and Table~\ref{Tab:directional spillovers quantile 0.95}, respectively.Compared to the risk spillover under the conditional mean and median, there is an excess price risk spillover in the extreme case, with $TCI$ of 41.2\% for the extreme lower quantile and $TCI$ of 38.7\% for the extreme upper quantile, which significantly exceeds the total risk spillover at the mean and median levels. It suggests that under extreme conditions, the portion of forecast error variance from agricultural futures themselves becomes smaller and the linkages between agricultural futures become stronger; this finding is consistent with previous findings that the dependence of agricultural markets is greater in the extreme quantiles \citep{FJ-Jiang-Su-Todorova-Roca-2016-JFuturesMark}.

In terms of the net spillover risk index, comparing the conditional mean and median cases, the risk roles of each agricultural futures as well as the intensity of risk transfer changes under extreme conditions. For the extreme lower quantile, which considers extreme downside risk, CZCE Strong Wheat shifts to become the top risk transmitter, followed by CBOT Soybean, CZCE Late Indica Rice, and then CBOT Corn, while CBOT Wheat shifts from a risk transmitter to a risk receiver, and DCE Corn shifts from a risk receiver to a risk transmitter. According to the table, it can be understood that in the case of extreme downward price movements, China's listed agricultural futures are no longer all in the downstream of risk transfer, which is closely related to China's status as a large agricultural producer and agricultural trade country, and also due to the government's introduction of food security policies. For the extreme upper quantile, which considers extreme upside risk, CZCE Wheat WH is also the top risk transmitter, followed by CZCE Late Indica Rice, while US-listed agricultural futures have all shifted from being risk transmitters to risk receivers, which is related to the different price limit systems of futures trading in both countries. In conclusion, China-listed agricultural futures take more initiative in extreme situations.

\begin{table}[thp]
  \renewcommand\arraystretch{1}
  \renewcommand\tabcolsep{2pt} 
   \centering
  \caption{Static risk spillover at the conditional median level}
  \begin{tabular}{lccccccccccccccc}
    \toprule
        & \multicolumn{4}{c}{CBOT} &   \multicolumn{5}{c}{CZCE} &   \multicolumn{2}{c}{DCE} & \multirow{2}{*}{FROM} \\
        \cmidrule(lr){2-5}  \cmidrule(lr){6-10} \cmidrule(lr){11-12}
        & Corn & RiceR  & Soybean &  Wheat & WheatWH & RiceE  & WheatPM & RiceJ  & 
        RiceLI  & Soybean & Corn & \\ 
    \midrule
    CBOT Corn & 56.5 & 1.8 & 17.7 & 18.3 & 1.0 & 0.6 & 1.2 & 0.8 & 0.7 & 0.7 & 0.5 & 43.5 \\ 
    CBOT RiceR  & 2.4 & 85.0 & 2.8 & 2.8 & 1.1 & 1.0 & 0.8 & 0.9 & 0.9 & 1.1 & 1.3 & 15 \\ 
    CBOT Soybean & 19.4 & 2.1 & 62.4 & 8.6 & 1.3 & 0.9 & 1.1 & 1.6 & 0.8 & 1.0 & 0.8 & 37.6 \\ 
    CBOT Wheat & 20.3 & 2.0 & 8.7 & 62.9 & 1.0 & 1.1 & 0.7 & 0.8 & 0.6 & 0.9 & 1.0 & 37.1 \\ 
    CZCE WheatWH & 1.2 & 0.8 & 1.3 & 1.4 & 88.4 & 1.2 & 0.8 & 0.9 & 1.7 & 1.1 & 1.1 & 11.6 \\ 
    CZCE Early Rice & 1.2 & 0.9 & 1.5 & 1.4 & 1.8 & 83.5 & 2.1 & 2.5 & 2.6 & 1.5 & 0.9 & 16.5 \\ 
    CZCE WheatPM & 1.9 & 1.2 & 1.5 & 1.1 & 0.9 & 2.1 & 85.3 & 2.2 & 2.1 & 1.0 & 0.8 & 14.7 \\ 
    CZCE RiceJ & 1.4 & 0.8 & 1.8 & 1.1 & 1.8 & 2.2 & 2.1 & 85.2 & 1.2 & 1.5 & 1.0 & 14.8 \\ 
    CZCE RiceLI & 0.7 & 0.5 & 0.7 & 0.8 & 1.4 & 1.5 & 1.3 & 0.7 & 90.0 & 1.2 & 1.1 & 10 \\ 
    DCE Soybean & 2.4 & 1.2 & 4.9 & 2.5 & 1.5 & 1.5 & 1.3 & 1.5 & 1.2 & 78.3 & 3.7 & 21.7 \\ 
    DCE Corn & 2.5 & 1.5  & 2.4  & 2.3  & 1.3  & 0.9  & 0.8  & 1.7  & 1.1  & 3.8  & 81.8 & 18.2 \\ 
    TO      & 53.4 & 12.7 & 43.4 & 40.4 & 13.0 & 13.0 & 12.4 & 13.7 & 12.8 & 13.7 & 12.2 & \\ 
    NET & 9.9 & $-2.2$ & 5.9  & 3.3  & 1.5  & -3.5 & -2.3 & -1.2 & 2.8  & -8.0 & -6.1 & TCI \\ 
    DOM & 9.0 & 3.0  & 7.0  & 8.0  & 6.0  & 1.0  & 1.0  & 4.0  & 6.0  & 1.0  & 0.0 & 21.9 \\ 
    \bottomrule
  \end{tabular}
  \begin{tablenotes}   
        \footnotesize        
        \item[1] Note: $\tau=0.5$
      \end{tablenotes}           
  \label{Tab:directional spillovers quantile 0.5}
\end{table}

\begin{table}[thp]
  \renewcommand\arraystretch{1}
  \renewcommand\tabcolsep{2pt} 
   \centering
  \caption{Static risk spillover in the extreme lower quantile}
  \begin{tabular}{lccccccccccccccc}
    \toprule
        & \multicolumn{4}{c}{CBOT} &   \multicolumn{5}{c}{CZCE} &   \multicolumn{2}{c}{DCE} & \multirow{2}{*}{FROM} \\
        \cmidrule(lr){2-5}  \cmidrule(lr){6-10} \cmidrule(lr){11-12}
        & Corn & RiceR  & Soybean &  Wheat & WheatWH & RiceE  & WheatPM & RiceJ  & 
        RiceLI  & Soybean & Corn & \\ 
    \midrule
   CBOT Corn & 49.6 & 2.4 & 14.8 & 13.1 & 4.2 & 2.7 & 2.3 & 3.3 & 3.0 & 1.4 & 3.1 & 50.4 \\ 
   CBOT RiceR & 3.0 & 63.6 & 6.1 & 2.8 & 3.7 & 2.5 & 2.9 & 3.3 & 5.2 & 2.0 & 4.9 & 36.4 \\ 
   CBOT Soybean & 14.4 & 4.4 & 49.2 & 7.3 & 4.6 & 3.6 & 2.5 & 3.6 & 3.7 & 2.4 & 4.5 & 50.8 \\ 
   CBOT Wheat & 12.9 & 3.8 & 9.1 & 48.8 & 5.0 & 2.3 & 3.4 & 3.5 & 4.8 & 1.6 & 4.8 & 51.2 \\ 
   CZCE WheatWH & 4.5 & 2.5 & 4.4 & 2.5 & 62.4 & 5.5 & 2.3 & 3.9 & 5.1 & 1.7 & 5.4 & 37.6 \\ 
   CZCE RiceE & 2.9 & 2.4 & 3.4 & 1.2 & 6.1 & 64.0 & 3.2 & 4.1 & 4.8 & 4.6 & 3.3 & 36 \\ 
   CZCE WheatPM & 2.8 & 3.7 & 2.8 & 2.7 & 3.1 & 3.7 & 64.4 & 3.7 & 5.9 & 2.0 & 5.2 & 35.6 \\ 
   CZCE RiceJ & 3.4 & 3.4 & 3.6 & 2.5 & 5.4 & 4.1 & 3.1 & 63.9 & 3.2 & 3.4 & 4.1 & 36.1 \\ 
   CZCE RiceLI & 3.0 & 3.3 & 3.6 & 3.4 & 5.2 & 4.3 & 4.2 & 2.5 & 64.1 & 2.4 & 4.0 & 35.9 \\ 
   DCE Soybean & 3.3 & 2.7 & 6.1 & 2.3 & 4.0 & 5.6 & 2.7 & 4.0 & 3.6 & 60.6 & 5.0 & 39.4 \\ 
   DCE Corn & 3.8 & 5.3 & 6.1 & 3.3 & 5.6 & 2.9 & 4.6 & 4.3 & 4.0 & 3.8 & 56.4 & 43.6 \\ 
   TO & 54.0 & 34.0 & 60.1 & 41.0 & 47.0 & 37.1 & 31.3 & 36.2 & 43.1 & 25.1 & 44.3 &  \\ 
   NET & 3.6 & -2.5 & 9.3 & -10.1 & 9.4 & 1.1 & -4.4 & 0.1 & 7.2 & -14.3 & 0.6 & TCI \\ 
   DOM & 7.0 & 5.0 & 6.0 & 2.0 & 9.0 & 5.0 & 2.0 & 4.0 & 7.0 & 0.0 & 4.0 & 41.2 \\ 
    \bottomrule
  \end{tabular}
  \begin{tablenotes}   
        \footnotesize        
        \item[1] Note: $\tau=0.05$
      \end{tablenotes}           
  \label{Tab:directional spillovers quantile 0.05}
\end{table}

\begin{table}[thp]
  \renewcommand\arraystretch{1}
  \renewcommand\tabcolsep{2pt} 
   \centering
  \caption{Static risk spillover in the extreme higher quantile}
  \begin{tabular}{lccccccccccccccc}
    \toprule
        & \multicolumn{4}{c}{CBOT} &   \multicolumn{5}{c}{CZCE} &   \multicolumn{2}{c}{DCE} & \multirow{2}{*}{FROM} \\
        \cmidrule(lr){2-5}  \cmidrule(lr){6-10} \cmidrule(lr){11-12}
        & Corn & RiceR  & Soybean &  Wheat & WheatWH & RiceE  & WheatPM & RiceJ  & 
        RiceLI  & Soybean & Corn & \\ 
    \midrule
   CBOT Corn & 53.2 & 3.1 & 12.7 & 10.2 & 4.4 & 2.5 & 2.2 & 3.1 & 2.9 & 2.6 & 3.2 & 46.8 \\ 
   CBOT RiceR & 2.8 & 65.4 & 3.5 & 4.5 & 3.0 & 3.0 & 3.0 & 3.2 & 4.3 & 4.0 & 3.4 & 34.6 \\ 
   CBOT Soybean & 12.9 & 3.4 & 51.5 & 6.3 & 5.9 & 3.8 & 2.3 & 4.2 & 4.0 & 3.0 & 2.8 & 48.5 \\ 
   CBOT Wheat & 10.0 & 4.8 & 6.4 & 54.7 & 5.4 & 3.5 & 2.7 & 3.7 & 3.7 & 2.3 & 2.7 & 45.3 \\ 
   CZCE WheatWH & 3.2 & 1.7 & 3.3 & 4.0 & 62.1 & 3.3 & 2.9 & 7.4 & 5.3 & 2.6 & 3.9 & 37.9 \\ 
   CZCE RiceE & 2.3 & 2.6 & 3.6 & 3.1 & 4.2 & 64.4 & 4.6 & 4.8 & 4.8 & 3.0 & 2.6 & 35.6 \\ 
   CZCE WheatPM & 1.6 & 2.7 & 2.1 & 1.9 & 3.7 & 4.3 & 68.4 & 3.8 & 5.4 & 3.4 & 2.7 & 31.6 \\ 
   CZCE RiceJ & 2.9 & 2.7 & 4.0 & 2.1 & 7.5 & 4.6 & 4.0 & 62.9 & 4.5 & 2.7 & 2.1 & 37.1 \\ 
   CZCE RiceLI & 1.7 & 2.4 & 3.1 & 2.6 & 4.8 & 3.1 & 4.6 & 4.1 & 66.5 & 2.8 & 4.2 & 33.5 \\ 
   DCE Soybean & 3.1 & 4.6 & 5.0 & 2.7 & 4.4 & 3.1 & 3.7 & 2.6 & 3.8 & 61.2 & 5.8 & 38.8 \\ 
   DCE Corn & 3.8 & 2.9 & 3.2 & 2.6 & 5.1 & 2.6 & 2.8 & 3.3 & 4.7 & 5.3 & 63.5 & 36.5 \\ 
   TO & 44.3 & 30.9 & 46.9 & 40.0 & 48.3 & 33.9 & 32.8 & 40.2 & 43.5 & 31.7 & 33.5 &  \\ 
   NET & -2.4 & -3.6 & -1.6 & -5.3 & 10.5 & -1.7 & 1.1 & 3.1 & 10.0 & -7.1 & -3.0 & TCI \\ 
   DOM & 3.0 & 3.0 & 4.0 & 2.0 & 9.0 & 5.0 & 8.0 & 6.0 & 10.0 & 1.0 & 3.0 & 38.7 \\ 
    \bottomrule
  \end{tabular}
  \begin{tablenotes}   
        \footnotesize        
        \item[1] Note: $\tau=0.95$
      \end{tablenotes}           
  \label{Tab:directional spillovers quantile 0.95}
\end{table}

\subsection{Quantile dynamic risk spillover effects}

\subsubsection{Dynamic total risk spillovers}

\begin{figure}[h!]
    \centering
    \includegraphics[width=0.6\linewidth]{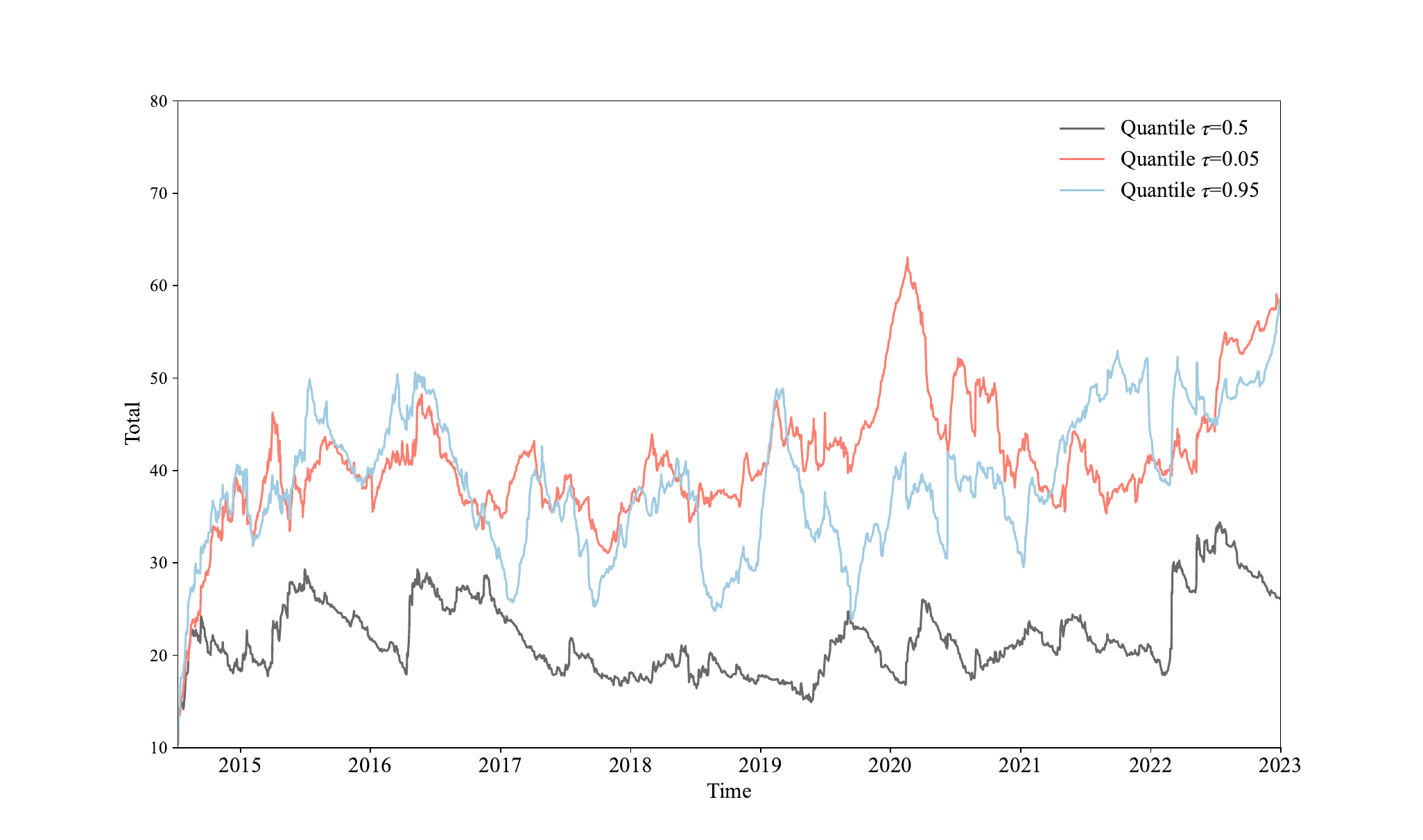}
    \vspace{-4mm}
    \caption{Total risk spillover under different quantiles}
    \label{Fig:AgroFutures_TVP_VAR_DY_quantileTOTAL}
\end{figure}

Fig.~\ref{Fig:AgroFutures_TVP_VAR_DY_quantileTOTAL} shows the total dynamic risk spillover among agricultural futures at the 0.5, 0.05, and 0.95 quantile levels. Observing the total dynamic risk spillover under the conditional median, it can be found that both the overall level and the shape of volatility are highly similar to Fig.~\ref{Fig:AgroFutures_TVP_VAR_DY_TOTAL}.

It is important to characterize the dynamic aggregate risk spillover under extreme market conditions. Based on the risk spillovers under extreme conditions presented in Fig.~\ref{Fig:AgroFutures_TVP_VAR_DY_quantileTOTAL}, it can be obtained that in the left and right tailed cases, the risk spillovers among agricultural futures are significantly higher than the level of total risk spillovers of the conditional mean or median, but the range of volatility is relatively small. The results suggest that agricultural futures are significantly more sensitive to extreme positive and negative shocks, with less volatility in the overall risk spillover, but are still affected by extreme events, such as the COVID-19 pandemic and the Russia-Ukraine conflict.

In general, as the financial attributes of agricultural commodities continue to grow, when the economy is in a period of turbulence or high uncertainty, the actual supply and demand risks are transmitted. In addition, all kinds of negative information and investor sentiments can spread rapidly, which leads to an increase in the correlation between agricultural commodity futures and an increase in the level of risk spillover, and thus the risk spillover between agricultural commodity futures in extreme cases needs to be paid more attention to. The impact of the global financial cycle and economic uncertainty on risk spillovers among agricultural futures is becoming increasingly critical.

\subsubsection{Dynamic directional risk spillovers}

\begin{figure}[!h]
    \centering
    \includegraphics[width=0.975\linewidth]{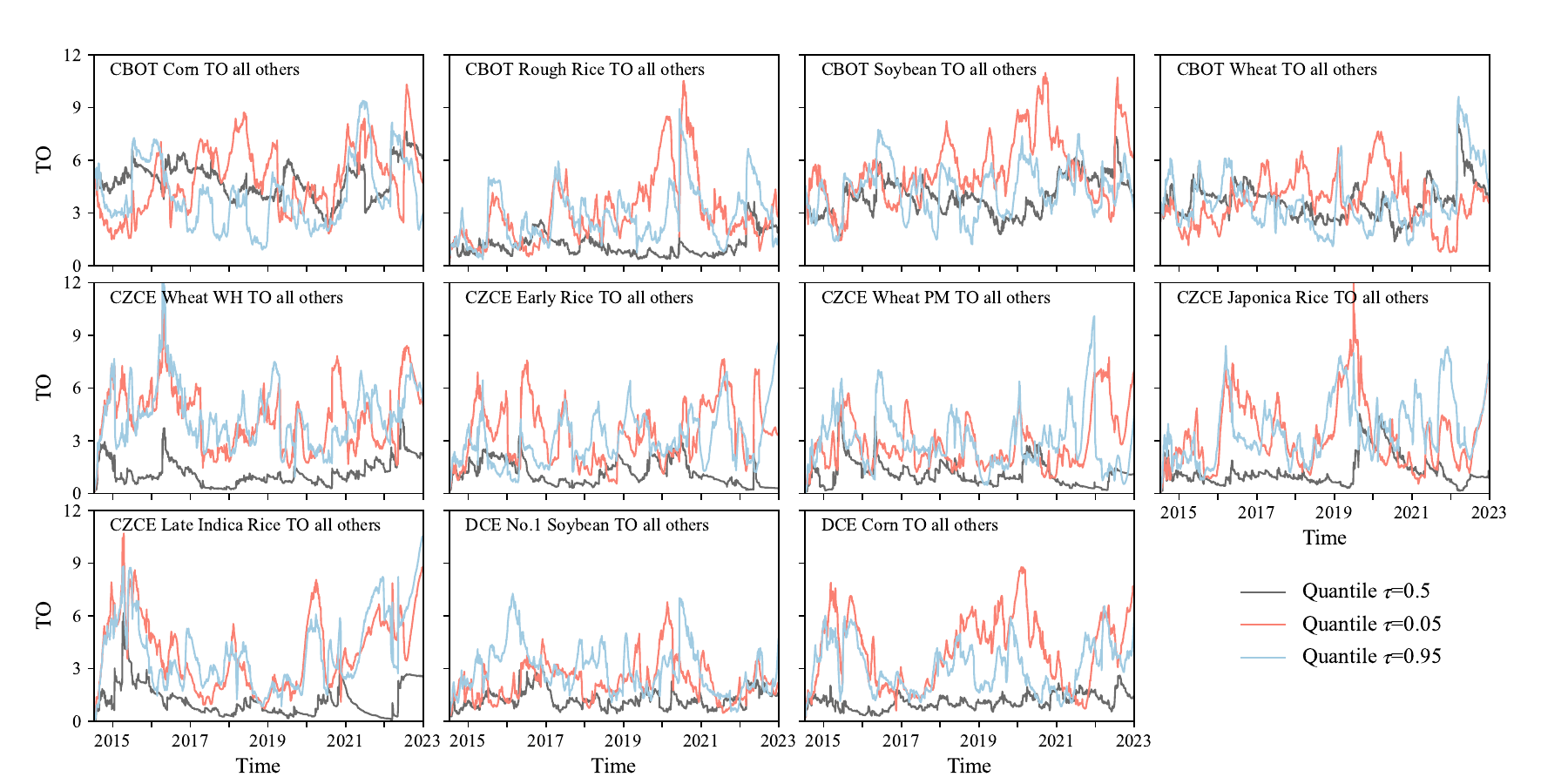}
    \caption{$TO$ spillover under different quantiles}
    \label{Fig:AgroFutures_TVP_VAR_DY_quantileTO}
\end{figure}

Fig.~\ref{Fig:AgroFutures_TVP_VAR_DY_quantileTO} shows the dynamic course of the directional spillover index $TO$ for 11 agricultural futures at the conditional median as well as the extreme lower and upper tails. The results obtained at the median level are consistent with those under the conditional mean, with the highest levels of risk spillover during the sample period for CBOT Corn, CBOT Soybean, and CBOT Wheat.

At the extreme, the level of $TO$ risk spillover for each agricultural futures is significantly larger than under the conditional mean or conditional median, with the spillover index reaching over 6\% at one point in time. Comparing the $TO$ under conditional mean and conditional median, CBOT Corn and soybean remain the largest risk spillover agricultural futures under extreme downside risk, but CBOT Wheat is no longer the strong risk spillover transmitter, while CZCE Wheat WH and CZCE Late Indica Rice show higher levels of risk spillover, and the periods of strong risk spillover for CZCE Wheat WH and CZCE Late Indica Rice are mainly concentrated in the periods of 2015–2016 and 2020–2022, i.e., the period of China's stock market crash and the period of pandemic recurrence, suggesting that the risk spillovers of both have certain geographical characteristics. In addition, in the extreme downside risk scenario, risk spillover  across agricultural futures is more intense when economic uncertainty rises. In contrast, the level of the $TO$ risk spillover index is smaller in the extreme upside scenario than in the extreme downside scenario.

\begin{figure}[!h]
    \centering
\includegraphics[width=0.975\linewidth]{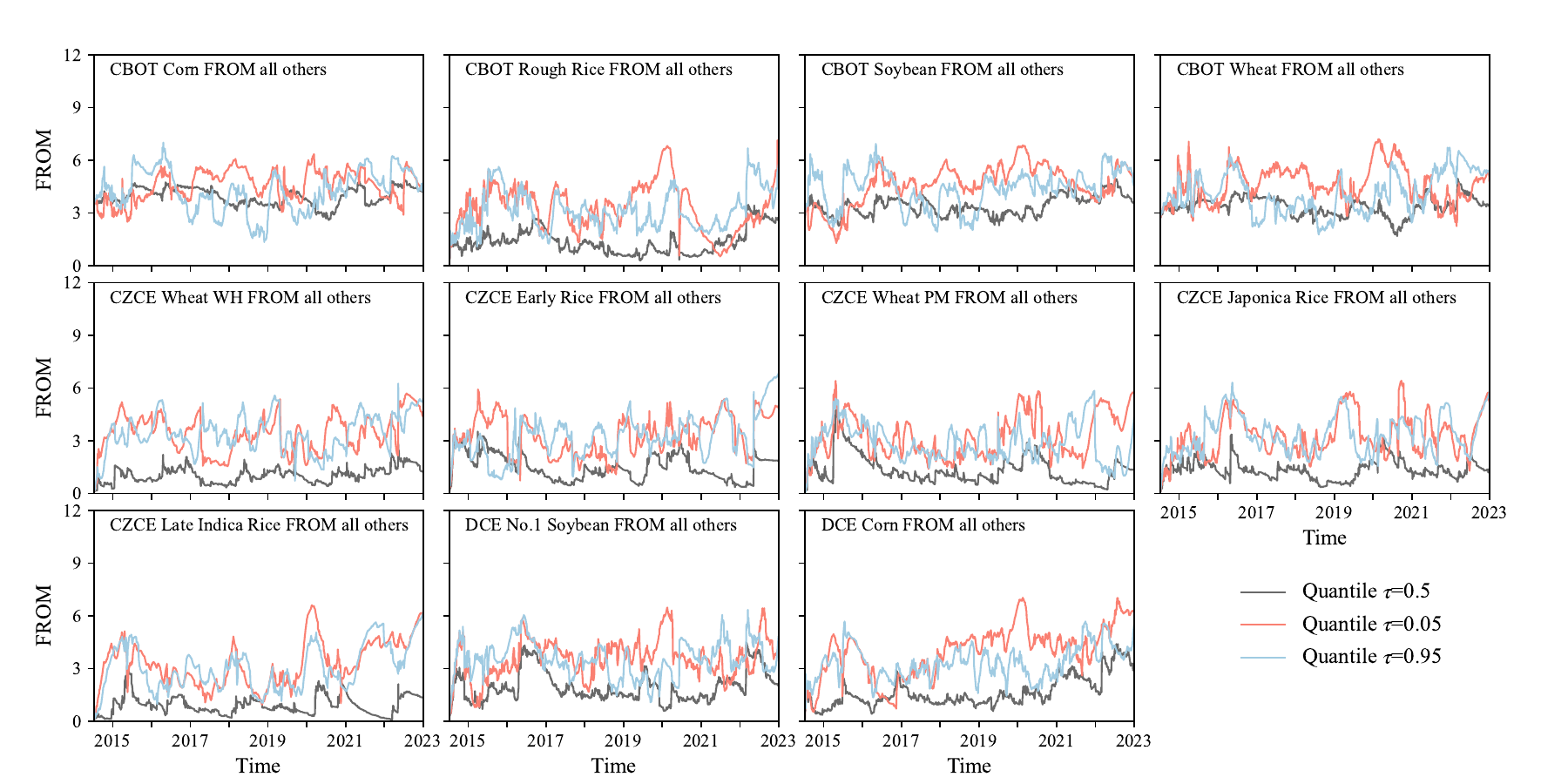}
    \caption{$FROM$ spillover under different quantiles}
    \label{Fig:AgroFutures_TVP_VAR_DY_quantileFROM}
\end{figure}

Fig.~\ref{Fig:AgroFutures_TVP_VAR_DY_quantileFROM} illustrates the dynamic course of the directional spillover indices $FROM$ for 11 agricultural futures under the conditional median and the extreme lower and upper quantile cases, respectively. The results under the conditional mean and the results under the conditional median are quite comparable, with CBOT corn, CBOT soybean, and CBOT wheat having the highest levels of risk premium.

In the extreme cases, the directional risk spillover $FROM$ is larger for all agricultural futures compared to the conditional mean and the conditional median. In contrast to the $TO$ index, the $FROM$ index has a relatively small range of volatility. Combined with $TO$ and $FROM$, it can be realized that the net risk exporter roles of CZCE Wheat WH and CZCE Late Indica Rice are mainly formed by transmitting a larger level of risk to other agricultural futures and receiving a smaller level of risk from the rest of the agricultural futures.

\subsubsection{Net risk spillovers}

\begin{figure}[!h]
    \centering
    \includegraphics[width=0.975\linewidth]{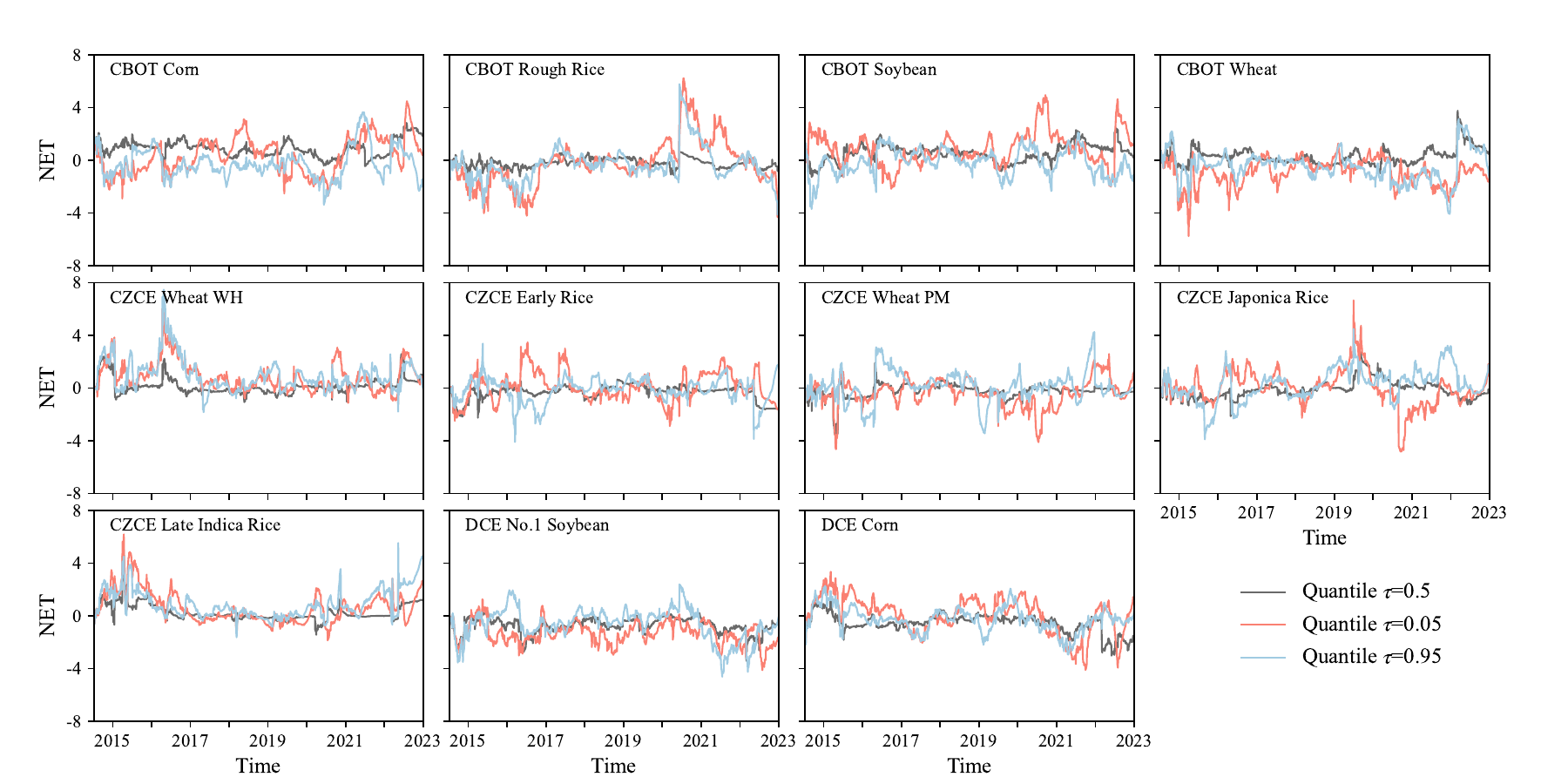}
    \caption{$NET$ spillover under different quantiles}
    \label{Fig:AgroFutures_TVP_VAR_DY_quantileNET}
\end{figure}

$Net$ risk spillover index can demonstrate the role of each agricultural commodity futures in the risk transfer process.
Fig.~\ref{Fig:AgroFutures_TVP_VAR_DY_quantileNET} presents the dynamic net spillover index among agricultural futures under the different quantiles. According to Fig.~\ref{Fig:AgroFutures_TVP_VAR_DY_quantileNET}, it is evident that there is a high degree of similarity between the conditional median and the conditional mean net spillover index results.
Unlike the results obtained for the conditional mean and conditional median net risk spillover, the determination of risk spillover roles is more difficult in extreme cases due to the fact that most agricultural futures roles tend to switch between risk transmitter and risk receiver during the overall sample period. Thus, in the extreme case, investors and policymakers need to pay constant attention to the risk spillover associated with the shock.

For the extreme downside case, the significant risk transmitters are CZCE Wheat WH, CZCE Late Indica Rice, and CBOT Soybean, and the risk receivers are CBOT Wheat and DCE No.1 Soybean, which is in line with the conclusions obtained from the previous quantile static risk spillover tables. For the extreme upside case, the only significant risk transmitters are CZCE Wheat WH and CZCE Late Indica Rice. The shift in the role of risk spillovers suggests that in extreme cases, CBOT-listed agricultural futures are no longer absolutely dominant; instead, some Chinese agricultural futures are upstream of the risk spillovers. The dominance of CZCE Wheat WH can be explained by the fact that China is a large producer as well as exporter of wheat, and CZCE Late Indica Rice is a risk transmitter probably because it is at the top of the domestic risk premium in China, linked to the Chinese financial markets as well as to policies. This is supported by the fact that the CZCE Late Indica Rice $NET$ spillover index under extreme risk remains large after the 2015 Chinese stock market crash and rises sharply after 2020 and 2022 after China's epidemic policies.

\subsection{Quantile risk spillover network and minimum spanning tree}

\subsubsection{Risk spillover network}

\begin{figure}[h!]
    \centering
    \includegraphics[width=0.32\linewidth]{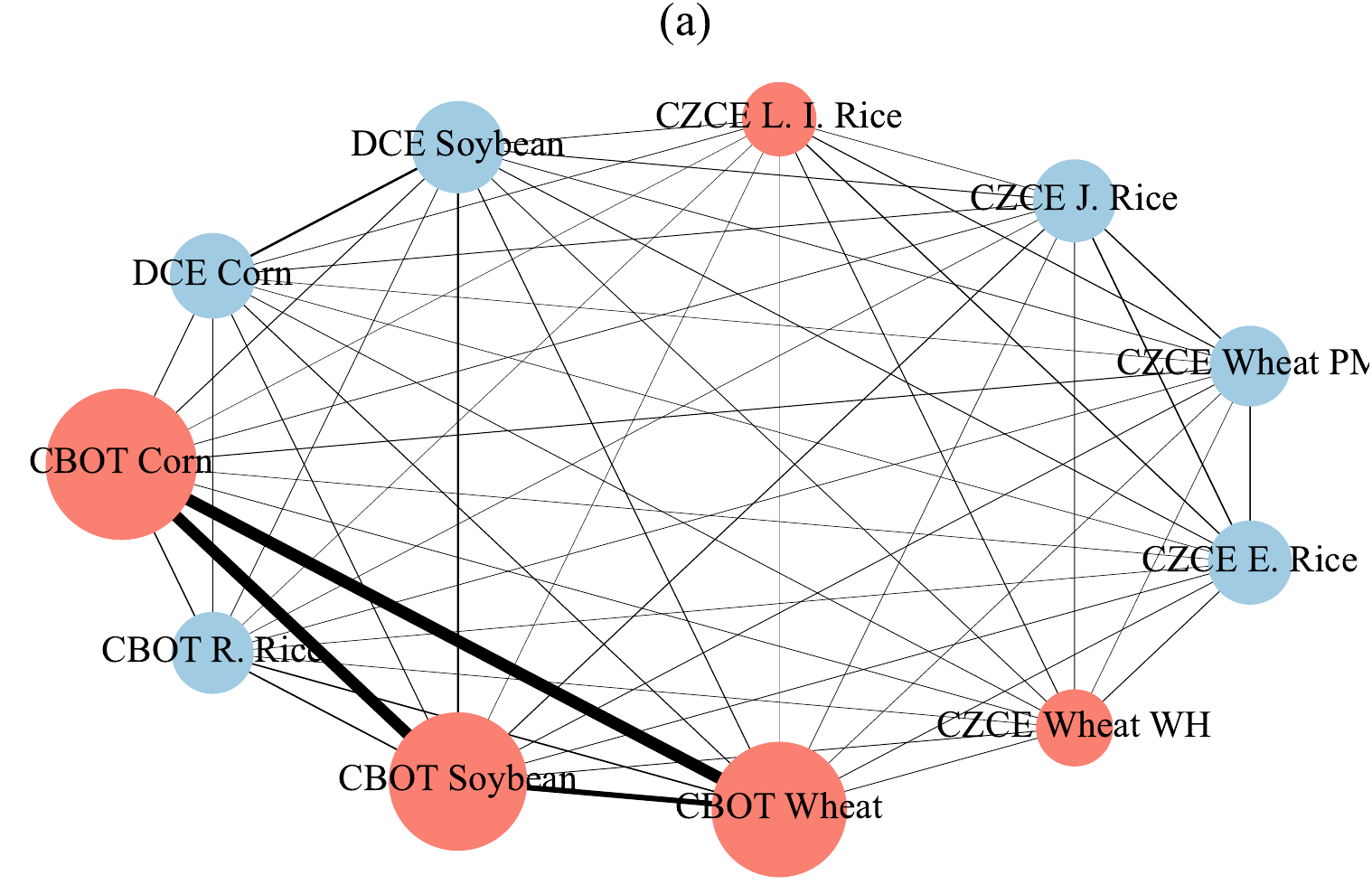}
    \includegraphics[width=0.32\linewidth]{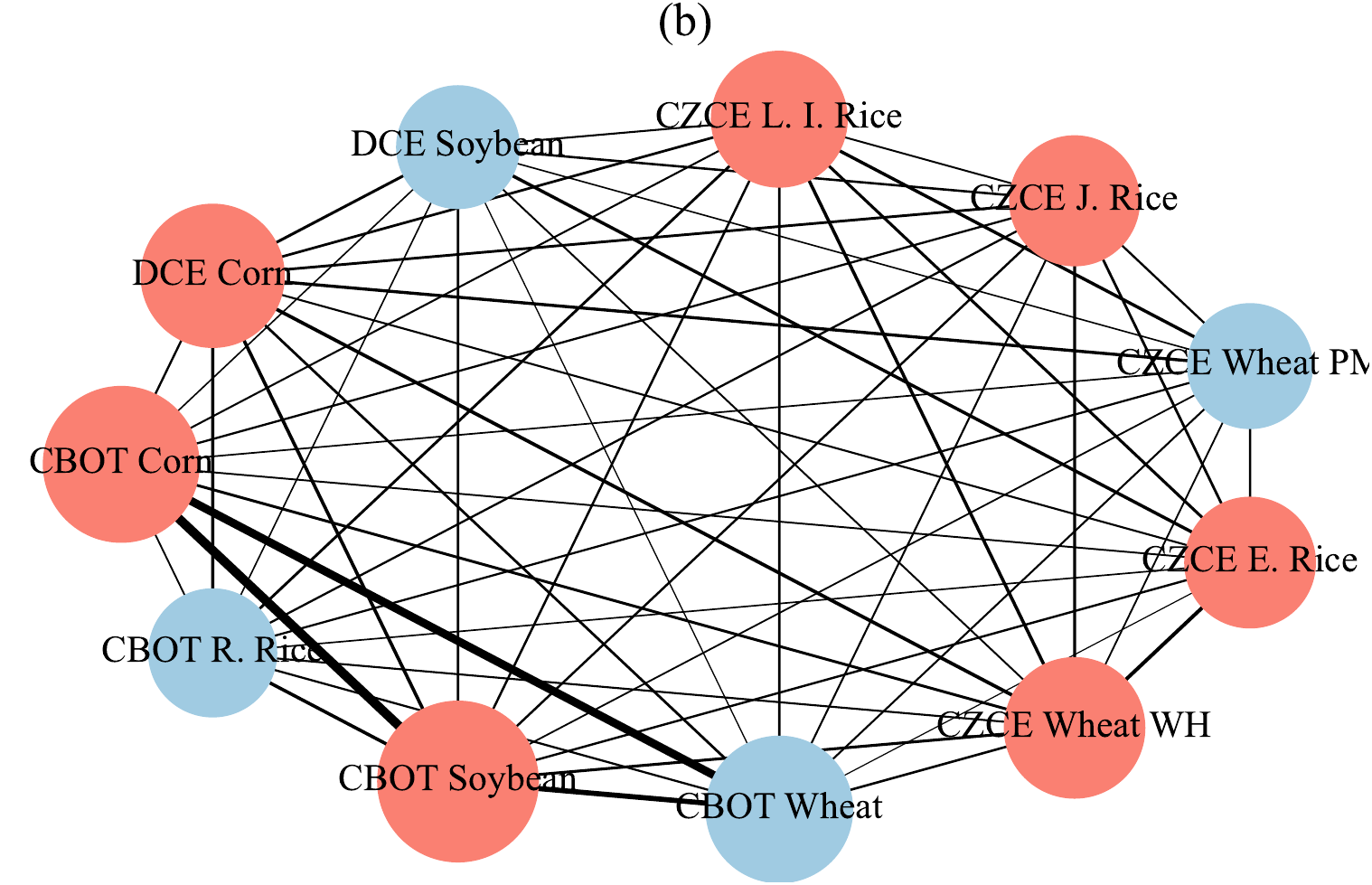}
    \includegraphics[width=0.32\linewidth]{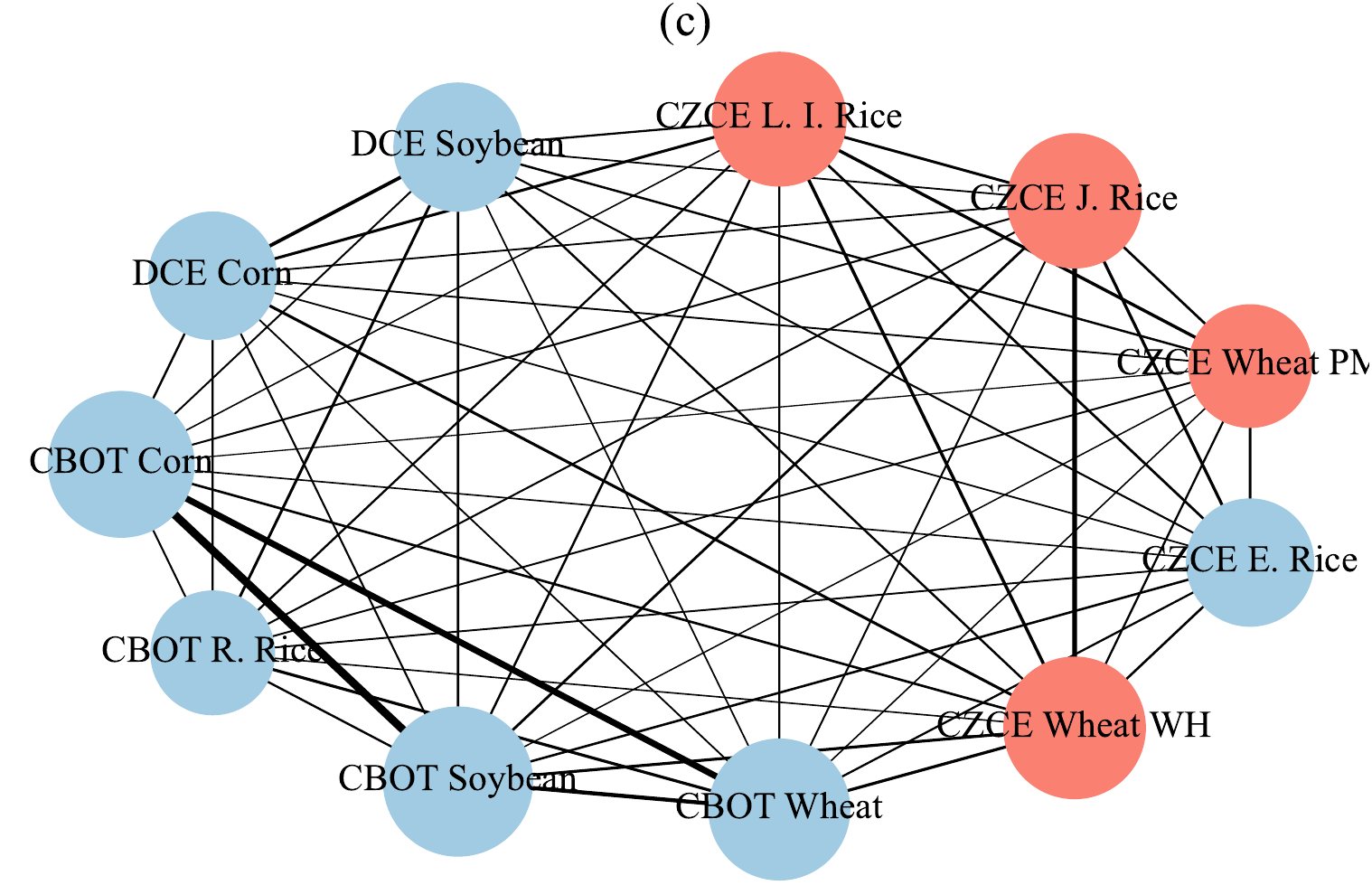}\\
    \includegraphics[width=0.32\linewidth]{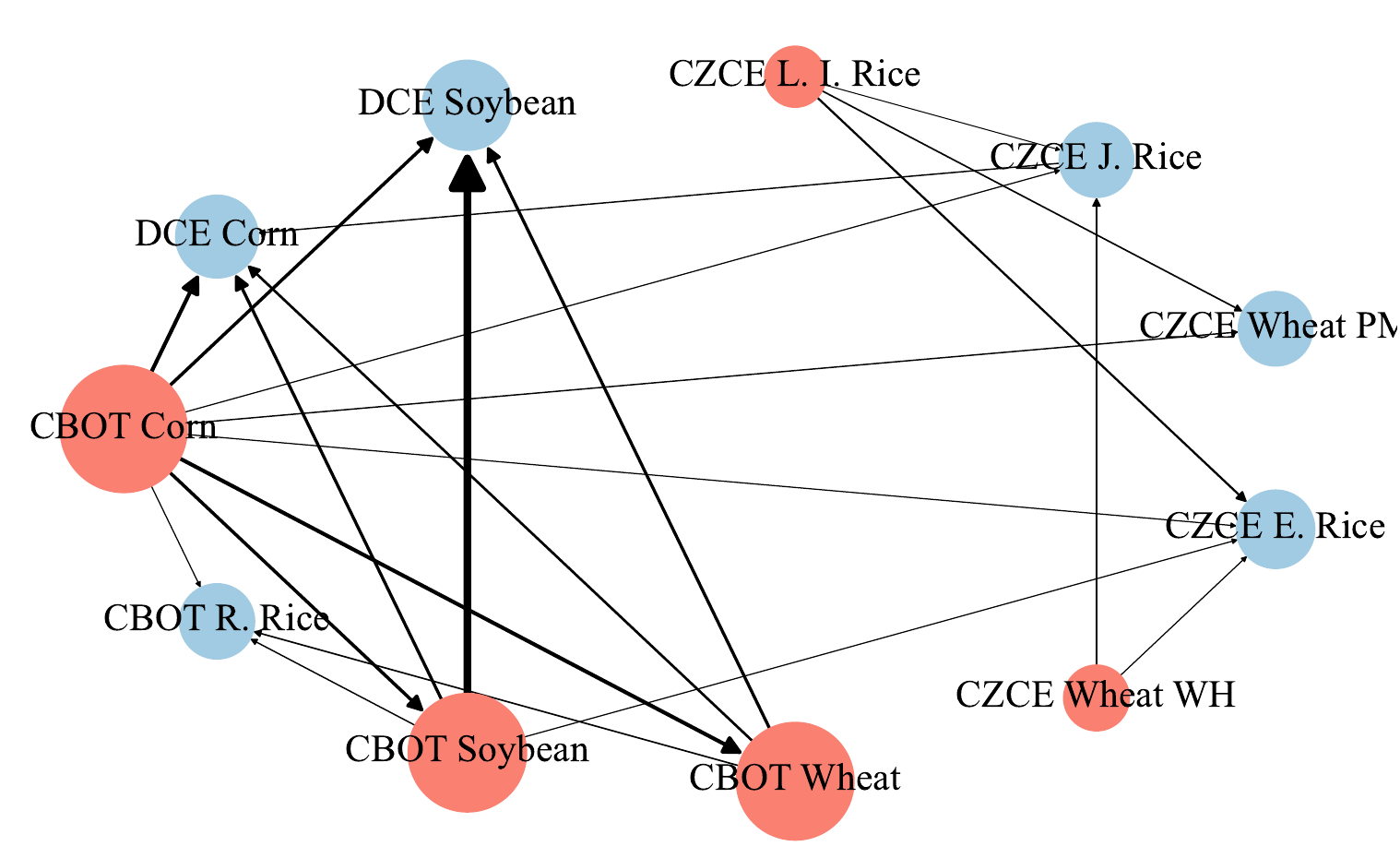}
    \includegraphics[width=0.32\linewidth]{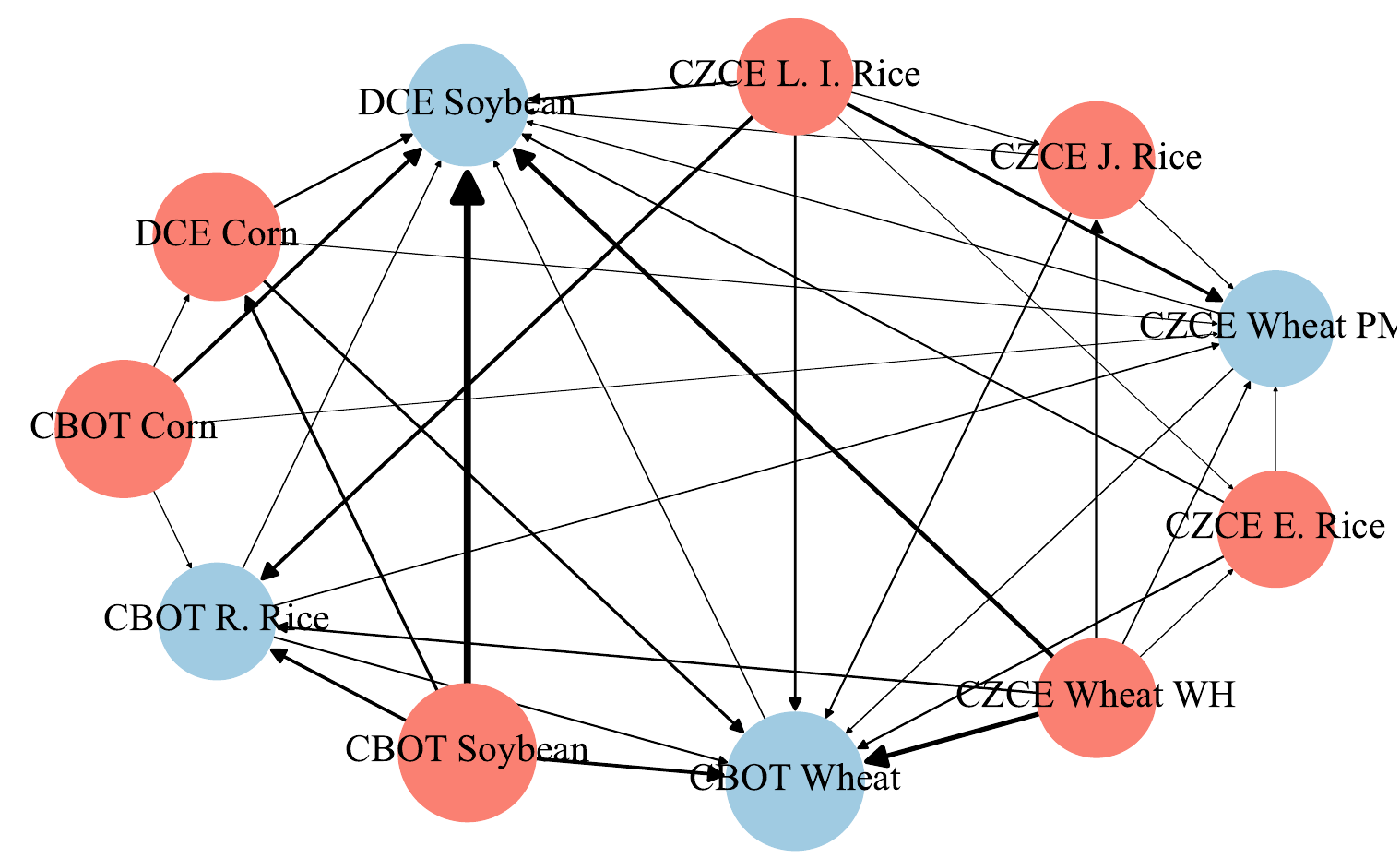}
    \includegraphics[width=0.32\linewidth]{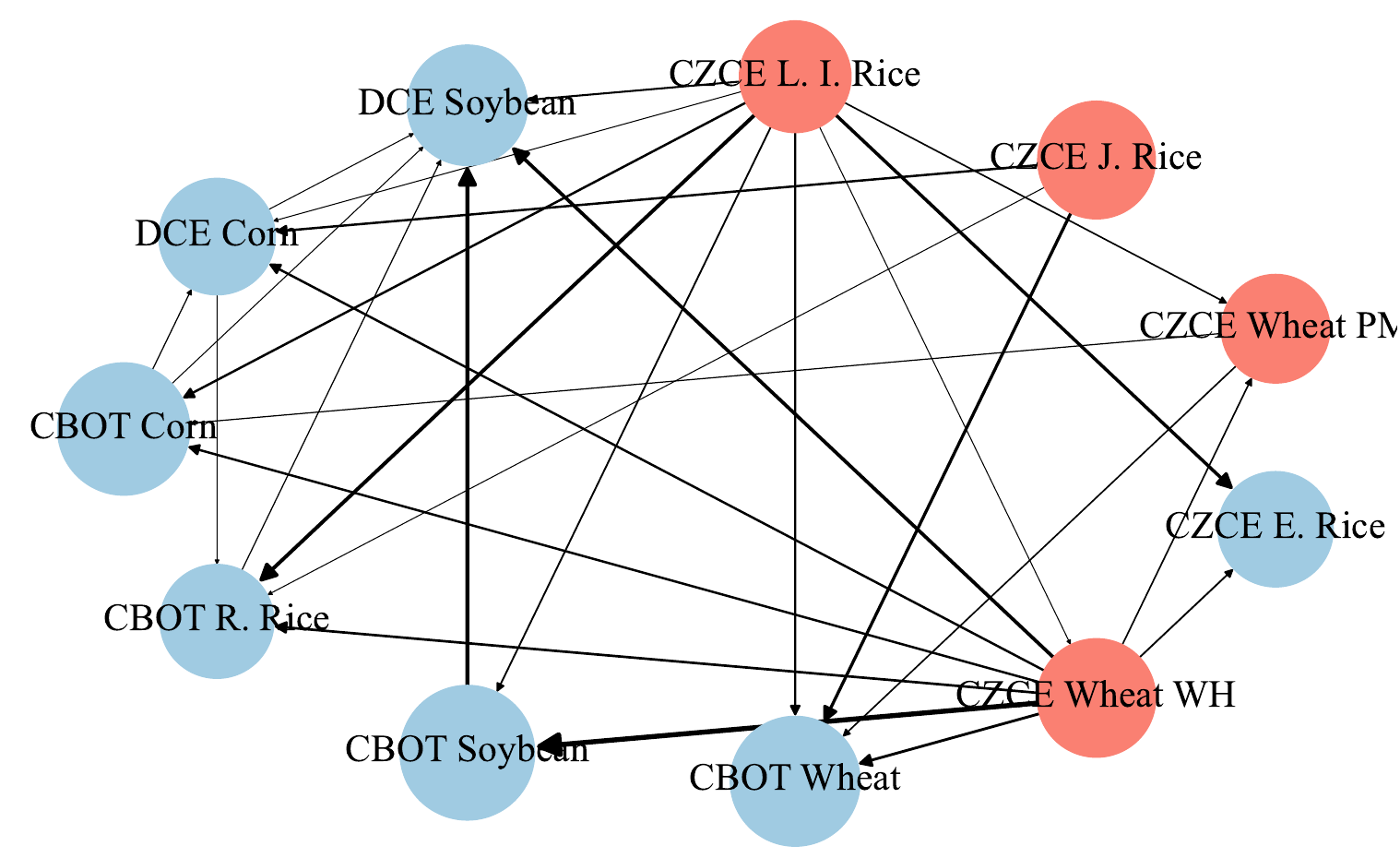}
    \caption{Spillover correlation network and net risk spillover at the conditional median level; Spillover correlation network and net risk spillover in the extreme lower quantile; Spillover correlation network and net risk spillover in the extreme upper quantile}
    \label{Fig:AgroFutures_TVP_VAR_DY_networks}
\end{figure}

Risk spillover networks can provide a more intuitive picture of the results of risk spillovers among agricultural futures, and this section provides visualization results by constructing correlation networks as well as net risk spillover networks under different quantiles.
The correlation network and the net spillover network under quantiles are constructed in the same way as the network under conditional means. Fig.~\ref{Fig:AgroFutures_TVP_VAR_DY_networks} shows the correlation network and the total net risk premium network of the  risk premium among agricultural futures under the conditional median, extreme downside, and extreme upside scenarios, respectively.

The conditional median agricultural futures risk spillover shown in Fig.~\ref{Fig:AgroFutures_TVP_VAR_DY_networks}(a) remains consistent with that under the conditional mean, but the net spillover risk network nodes are more connected, i.e., the level of the net spillover between agricultural futures is higher under the conditional median than under the conditional mean. 
In contrast, Fig.~\ref{Fig:AgroFutures_TVP_VAR_DY_networks}(b) and Fig.~\ref{Fig:AgroFutures_TVP_VAR_DY_networks}(c), which consider the extreme risk scenario, deviate from the average scenario. Observing the nodes, it can be obtained that the average strength and out-strength of the nodes in the extreme cases are larger than those under the conditional mean and the conditional median, and the roles of nodes have some changes. For the extreme downside scenario, CBOT Wheat switches from risk transmitter to risk receiver, while CZCE Japonica Rice and CZCE Early Indica Rice switch from risk receiver to risk transmitter. For extreme upside scenarios, CBOT Corn, CBOT Soybean, and CBOT Wheat all switch from risk exporters to risk receivers. The change in the role of nodes suggests a shift in the dominance of risk spillovers from CBOT-listed agricultural futures to CZCE-listed agricultural futures.
The net risk spillover links between nodes in the extreme case also differ somewhat from the average as well. First, the number of net risk spillover links in the extreme case is higher than that in the conditional average and conditional median, suggesting that the linkages among agricultural futures become stronger in the extreme case. Second, the net risk spillover relationship between CBOT Corn, Soybean, and Wheat is no longer shown on the net risk spillover network, but it can be seen from the correlation network that the relationship is still relatively strong, suggesting that spillovers between the three are offset in extreme cases. In addition, the net risk spillover effect of CZCE Late Indica Rice and CZCE Wheat WH is enhanced.

\subsubsection{Minimum spanning tree}

\begin{figure}[!h]
    \centering
    \includegraphics[width=0.323\linewidth]{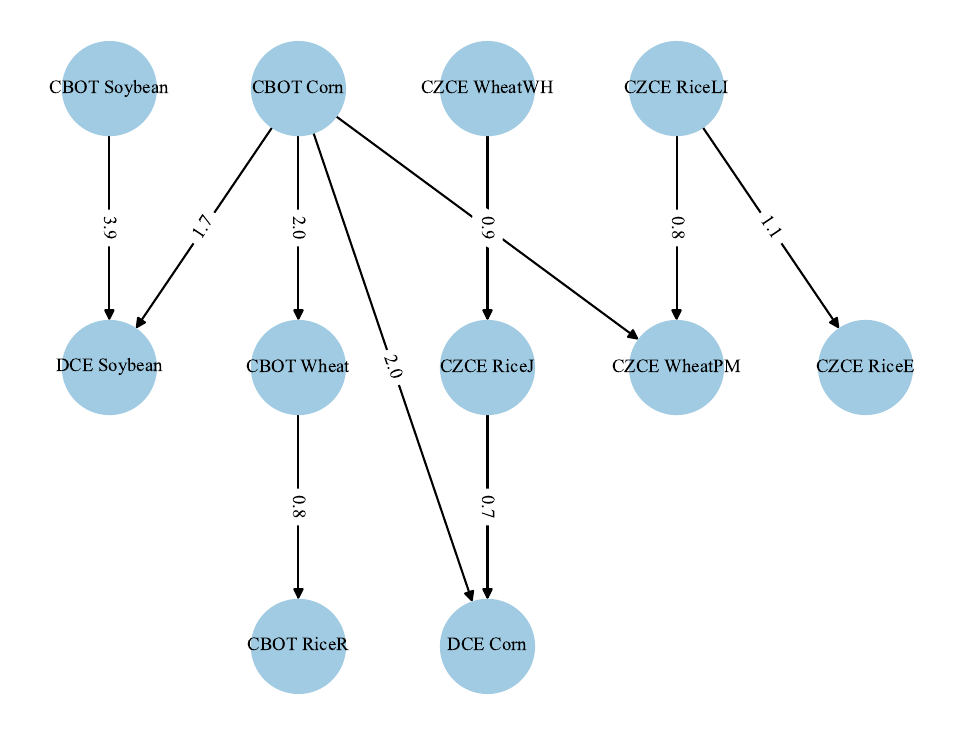}
    \includegraphics[width=0.323\linewidth]{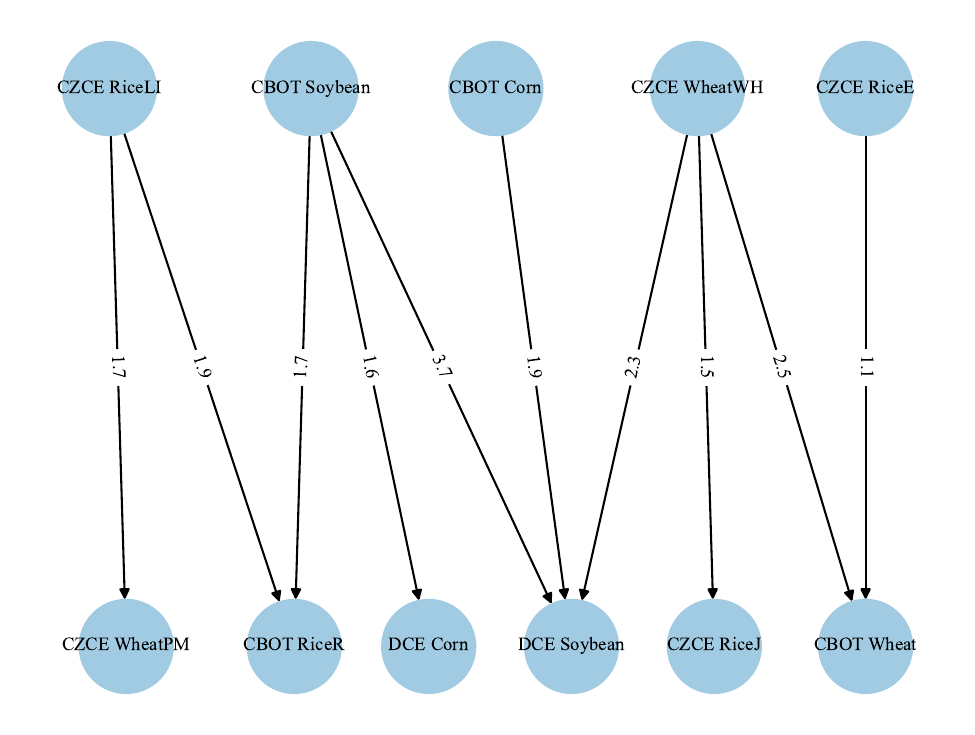}
    \includegraphics[width=0.323\linewidth]{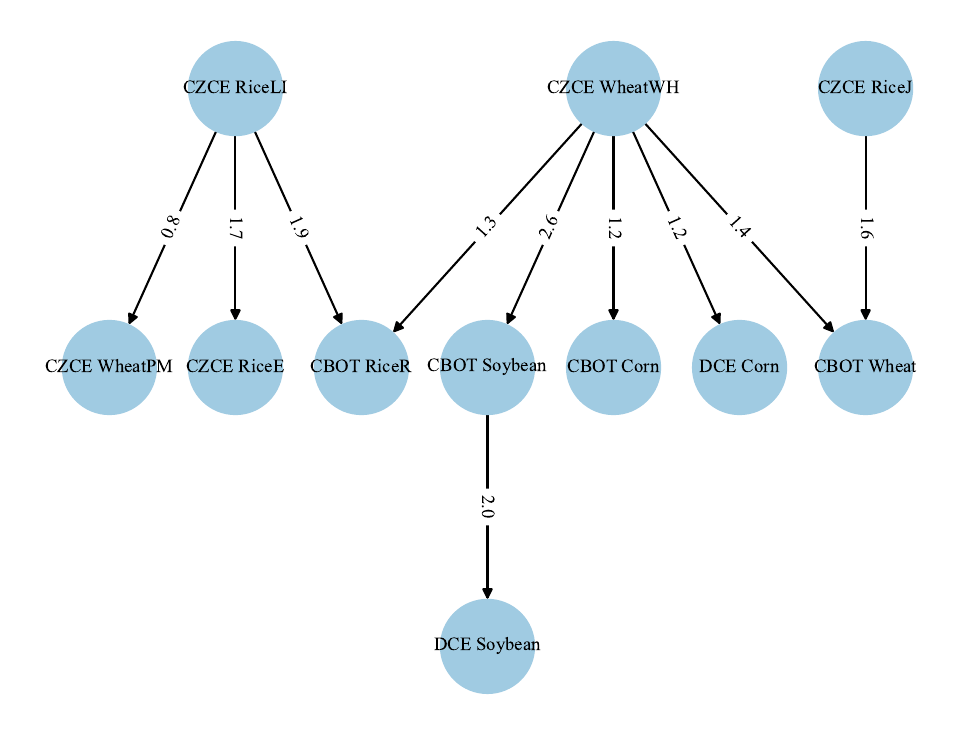}
    \vspace{-2mm}
    \caption{Minimum spanning tree under the conditional median level (left), in the extreme lower quantile (middle), and in the extreme upper quantile (right).}
    \label{Fig:AgroFutures_TVP_VAR_DY_quantileMST}
\end{figure}

In the quantile framework, the study also analyzes the risk spillover paths using the minimum spanning tree method, constructed in the same way as the minimum spanning tree under the conditional mean.Fig.~ \ref{Fig:AgroFutures_TVP_VAR_DY_quantileMST} presents the minimum spanning tree plot under conditional median, extreme downside risk, and extreme upside risk, with the node arrows representing the direction of risk spillovers and the numbers on the connections representing the magnitude of the pairwise net spillover risk between the two agricultural futures.

Comparison of Fig.~\ref{Fig:AgroFutures_TVP_VAR_DY_quantileMST}(a) with Fig.~\ref{Fig:AgroFutures_TVP_VAR_DY_MST} reveals that the minimum spanning tree graphs under the conditional median no longer have a very clear top-to-bottom hierarchical structure, and monotonicity centrality is significantly weaker than the conditional median case.
Considering the extreme case, the minimum spanning tree for extreme downside risk has some symmetry with the minimum spanning tree for extreme upside risk. In the extreme downside case, the risk spill center agricultural futures are CZCE Wheat WH, CBOT Soybean, and CZCE Late Indica Rice, and in the extreme upside case, the risk spill center agricultural futures are CZCE Wheat WH and CZCE Late Indica Rice. Meanwhile, together with Fig.~\ref{Fig:AgroFutures_TVP_VAR_DY_networks}(b) and Fig.~\ref{Fig:AgroFutures_TVP_VAR_DY_networks}(c), it can be found that the risk spillover targets of the center agricultural futures have some differences in types, with CZCE Late Indica Rice having risk spillover mainly to rice-based agricultural futures and CZCE Wheat WH having risk spillover mainly to wheat and soybeans.

\subsection{Robustness test}

\begin{figure}[!h]
    \centering
    \includegraphics[width=0.43\linewidth]{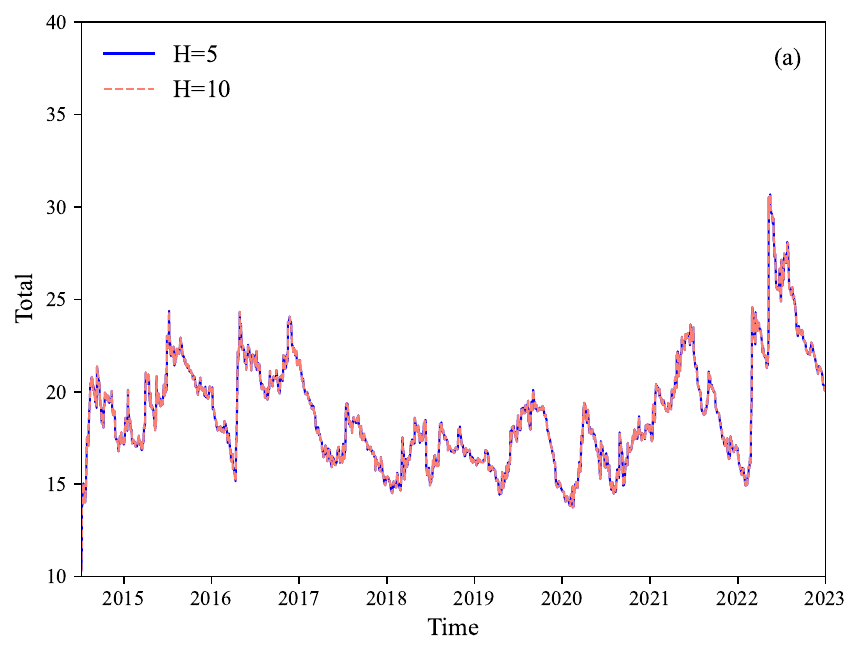}
    \includegraphics[width=0.43\linewidth]{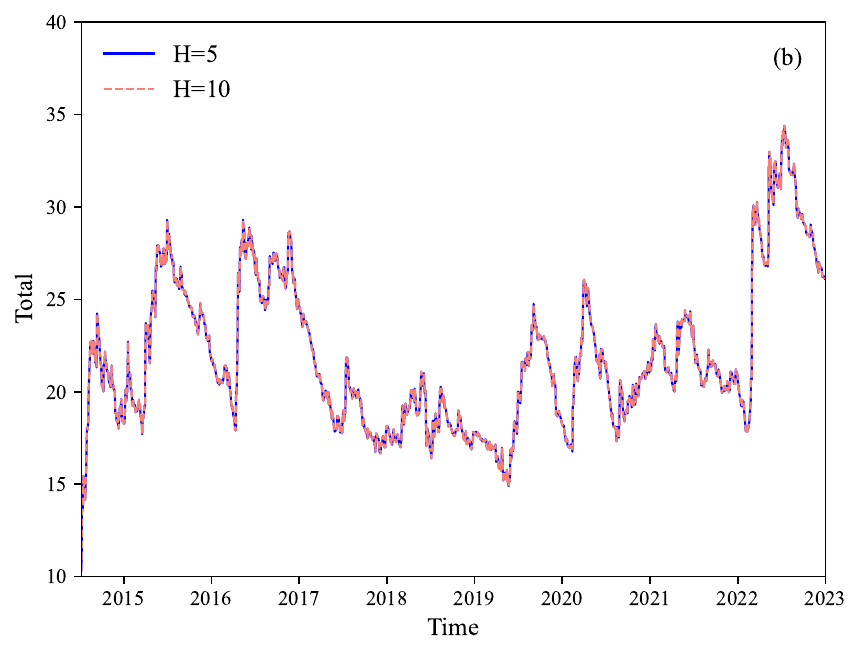}
    \includegraphics[width=0.43\linewidth]{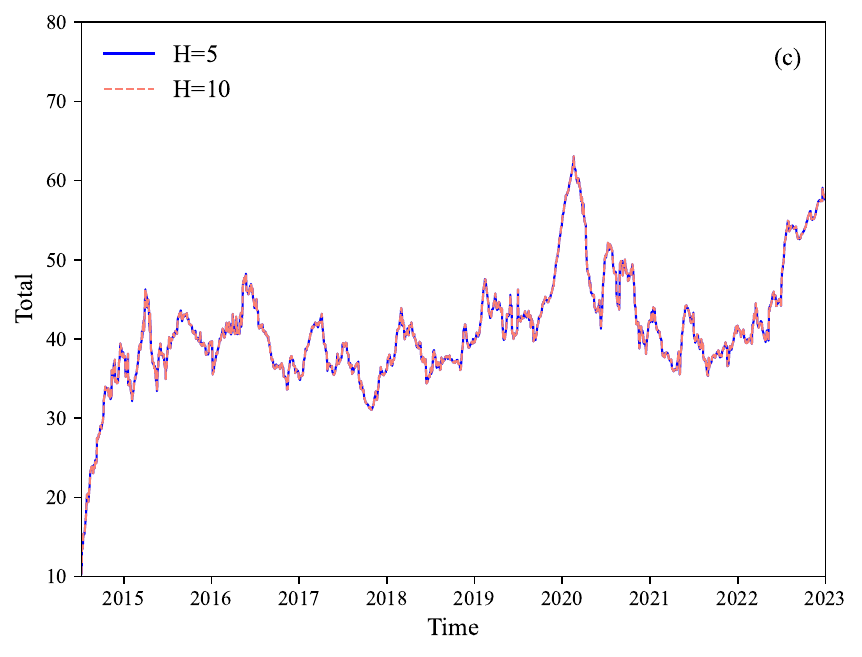}
    \includegraphics[width=0.43\linewidth]{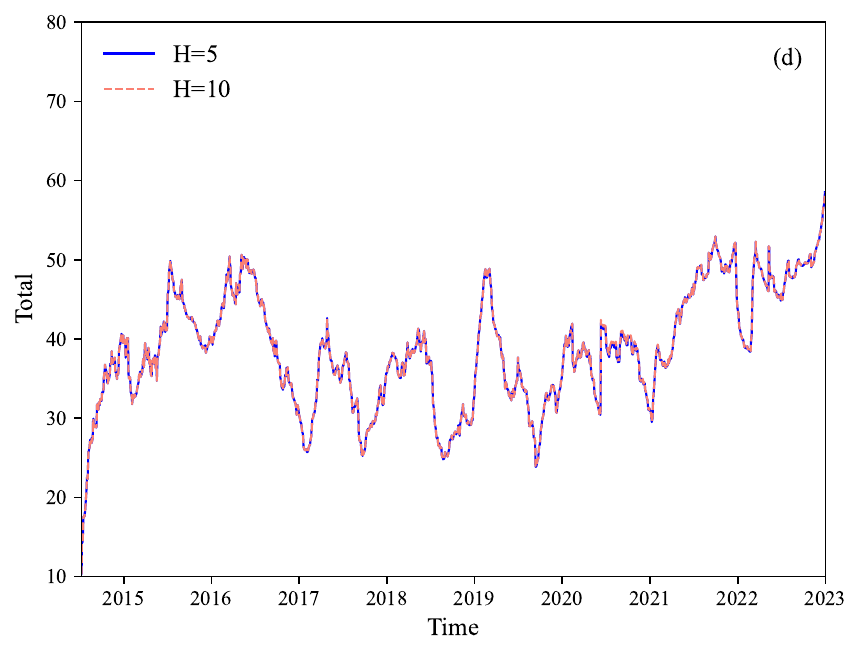}
    \caption{Robustness test}
    \label{Fig:AgroFutures_TVP_VAR_DY_TOTAL_Robustness}
\end{figure}

To further ensure the reliability of the empirical results, our paper changes the forecasting step from 5 periods forward to 10 periods forward in the case of the conditional mean and quantiles and obtains the results of the total dynamic spillover index as shown in Fig.~\ref{Fig:AgroFutures_TVP_VAR_DY_TOTAL_Robustness}. Fig.~\ref{Fig:AgroFutures_TVP_VAR_DY_TOTAL_Robustness}(a) shows the robustness test under conditional mean, Fig.~\ref{Fig:AgroFutures_TVP_VAR_DY_TOTAL_Robustness}(b) shows the robustness test under conditional median , Fig.~\ref{Fig:AgroFutures_TVP_VAR_DY_TOTAL_Robustness}(c) and (d) are robustness tests under extreme downside and extreme upside. From the figure, we can see that the trend of risk spillover effects for different forecasting steps is basically the same, and it can be proved that the results are not sensitive to the change of forecasting steps and are robust. 

\section{Conclusions}
\label{S1:Conclude}

It is impossible to overstate the value of agricultural products to humanity, and as average unit yields rise as a result of scientific and technological advancement, agricultural products continue to be financialized, and the level of economic globalization rises, food crises caused by price volatility, rather than a lack of production, are gradually taking over as the primary cause of the current food crisis.
Therefore, our paper mainly considers price risk. Based on the TVP-VAR-DY model under conditional mean as well as quantile, our paper analyzes the risk spillover effect between 11 agricultural futures listed in China and the US from July 9, 2014, to December 31, 2022, and obtains the risk spillover results. Meanwhile, for the convenience of observation, the results of the risk spillover effect are network visualized in this paper.

Our main results suggest that risk correlations across agricultural futures are high, risk spillovers are stronger in the extreme case than in the general case, and the risk roles of each agricultural futures change across different scenarios. 
There is a high degree of similarity between the risk spillover results for the conditional mean case and the conditional median case. At the conditional mean as well as the conditional median, the main net transmitters of risk are CBOT Corn and CBOT Soybean, and the main net receivers of risk are DCE Corn and DCE No.1  Soybean. Unexpected events or increased economic uncertainty can lead to a significant increase in overall risk spillovers, mainly caused by the impact of uncertainty on agricultural exports and imports and deeper financialization. 
In both the conditional mean and conditional median cases, risk contagion is mainly focused on corn and soybeans from a type perspective, with rice playing an insignificant role in risk transmission and being the least affected. In addition, in the risk spillover process, China is in a weak position and the United States is in a strong position, which can be explained by the fact that the two countries are large importers and exporters of agricultural products, respectively. 
By visualizing the network with the minimum spanning tree, we can understand that, in general, CBOT corn is the most important player in the risk spillover and is located in the most upstream of the risk spillover.

Risk spillover results at the extremes are quite different from those under the conditional mean and conditional median. In extreme downside and upside scenarios, the main risk transmitters turn to CZCE Wheat WH and CZCE Late Indica Rice, and the main risk receivers are DCE No.1 Soybean and CBOT Wheat. 
From a dynamic perspective, the total risk spillover at the extremes is largely above the conditional mean and conditional median levels over the sample period, but the range of fluctuations is smaller than that. And in the extreme case, the role of agricultural futures in the risk spillover process is relatively ambiguous compared to the conditional mean or conditional median throughout the sample period, with each agricultural futures shifting between risk transmiter and receiver more frequently. 
Using the risk spillover network with the minimum spanning tree, we find that the center nodes of risk spillover are CZCE Wheat WH and CZCE Late Indica Rice, CZCE Wheat WH mainly spills risk to Soybean and Wheat, and CZCE Late Indica Rice mainly spills risk to Rice cereal class.

Based on the above findings, the insights gained from the study and the recommendations made are as follows.
For decision-makers, the main focus is to use the direction of price risk transfer to ensure the relative stability of agricultural commodity prices. First, there is a need to guard against agricultural commodity futures price risk due to sudden economic events. Second, on average, it is necessary to pay attention to the price fluctuations of CBOT Corn, Soybean, and Wheat, and in extreme cases, focus on CZCE Wheat WH as well as Late Indica Rice, considering their price fluctuations as risk spillover leading indicators. Third, on the basis of ensuring the supply and price stability of staple grains, i.e., rice and wheat, it is important to monitor the price risk of soybeans and corn, and consider diversified import routes to reduce the risk of one-way spillovers, as well as continuously supporting the development of agricultural science and technology to improve unit yields.
Investors can construct investment portfolios with reasonable risk avoidance, according to the conclusions of this paper. First, investors can customize their investment strategies according to the risk spillover relationship in different situations, and when the price of agricultural products located upstream of the risk spillover rises, they can invest in their risk contagion targets, thus constructing a better investment portfolio. Second, rice can be considered a risk-avoidance option in the average case due to its weak level of risk transmission and reception and its weak correlation with other agricultural products; in extreme cases, risk spillover needs to be continuously monitored.



\section*{Data availability}

The data set in this paper are sourced from the Wind database (https://www.wind.com.cn) and the website of the International Grains Council (https://www.igc.int).

%
\bibliography{Bib}

\begin{thebibliography}{40}
\expandafter\ifx\csname natexlab\endcsname\relax\def\natexlab#1{#1}\fi
\providecommand{\url}[1]{\texttt{#1}}
\providecommand{\href}[2]{#2}
\providecommand{\path}[1]{#1}
\providecommand{\DOIprefix}{doi:}
\providecommand{\ArXivprefix}{arXiv:}
\providecommand{\URLprefix}{URL: }
\providecommand{\Pubmedprefix}{pmid:}
\providecommand{\doi}[1]{\href{http://dx.doi.org/#1}{\path{#1}}}
\providecommand{\Pubmed}[1]{\href{pmid:#1}{\path{#1}}}
\providecommand{\bibinfo}[2]{#2}
\ifx\xfnm\relax \def\xfnm[#1]{\unskip,\space#1}\fi
\bibitem[{Adeleke and Awodumi(2022)}]{FJ-Adeleke-Awodumi-2022-JApplEcon}
\bibinfo{author}{Adeleke, M.A.}, \bibinfo{author}{Awodumi, O.B.},
  \bibinfo{year}{2022}.
\newblock \bibinfo{title}{Modelling time and frequency connectedness among
  energy, agricultural raw materials and food markets}.
\newblock \bibinfo{journal}{Journal of Applied Economics} \bibinfo{volume}{25},
  \bibinfo{pages}{644--662}.
\newblock \DOIprefix\doi{10.1080/15140326.2022.2056300}.
\bibitem[{Adrian and
  Brunnermeier(2016)}]{FJ-Adrian-Brunnermeier-2016-AmEconRev}
\bibinfo{author}{Adrian, T.}, \bibinfo{author}{Brunnermeier, M.K.},
  \bibinfo{year}{2016}.
\newblock \bibinfo{title}{{CoVaR}}.
\newblock \bibinfo{journal}{American Economic Review} \bibinfo{volume}{106},
  \bibinfo{pages}{1705--1741}.
\newblock \DOIprefix\doi{10.1257/aer.20120555}.
\bibitem[{A{\"{i}}t-Youcef(2019)}]{FJ-AitYoucef-2019-EconModel}
\bibinfo{author}{A{\"{i}}t-Youcef, C.}, \bibinfo{year}{2019}.
\newblock \bibinfo{title}{How index investment impacts commodities: A story
  about the financialization of agricultural commodities}.
\newblock \bibinfo{journal}{Economic Modelling} \bibinfo{volume}{80},
  \bibinfo{pages}{23--33}.
\newblock \DOIprefix\doi{10.1016/j.econmod.2018.04.007}.
\bibitem[{Antonakakis et~al.(2020)Antonakakis, Chatziantoniou and
  Gabauer}]{Antonakakis-Chatziantoniou-Gabauer-2020-RFM}
\bibinfo{author}{Antonakakis, N.}, \bibinfo{author}{Chatziantoniou, I.},
  \bibinfo{author}{Gabauer, D.}, \bibinfo{year}{2020}.
\newblock \bibinfo{title}{Refined measures of dynamic connectedness based on
  time-varying parameter vector autoregressions}.
\newblock \bibinfo{journal}{Journal of Risk and Financial Management}
  \bibinfo{volume}{13}, \bibinfo{pages}{84}.
\newblock \DOIprefix\doi{10.3390/jrfm13040084}.
\bibitem[{Apostolakis et~al.(2022)Apostolakis, Floros and
  Giannellis}]{FJ-Apostolakis-Floros-Giannellis-2022-IntRevEconFinanc}
\bibinfo{author}{Apostolakis, G.N.}, \bibinfo{author}{Floros, C.},
  \bibinfo{author}{Giannellis, N.}, \bibinfo{year}{2022}.
\newblock \bibinfo{title}{On bank return and volatility spillovers:
  {I}dentifying transmitters and receivers during crisis periods}.
\newblock \bibinfo{journal}{International Review of Economics \& Finance}
  \bibinfo{volume}{82}, \bibinfo{pages}{156--176}.
\newblock \DOIprefix\doi{10.1016/j.iref.2022.06.009}.
\bibitem[{Barbaglia et~al.(2020)Barbaglia, Croux and
  Wilms}]{FJ-Barbaglia-Croux-Wilms-2020-EnergyEcon}
\bibinfo{author}{Barbaglia, L.}, \bibinfo{author}{Croux, C.},
  \bibinfo{author}{Wilms, I.}, \bibinfo{year}{2020}.
\newblock \bibinfo{title}{Volatility spillovers in commodity markets: A large
  t-vector autoregressive approach}.
\newblock \bibinfo{journal}{Energy Economics} \bibinfo{volume}{85},
  \bibinfo{pages}{104555}.
\newblock \DOIprefix\doi{10.1016/j.eneco.2019.104555}.
\bibitem[{Basak and Pavlova(2016)}]{FJ-Basak-Pavlova-2016-JFinanc}
\bibinfo{author}{Basak, S.}, \bibinfo{author}{Pavlova, A.},
  \bibinfo{year}{2016}.
\newblock \bibinfo{title}{A model of financialization of commodities}.
\newblock \bibinfo{journal}{Journal of Finance} \bibinfo{volume}{71},
  \bibinfo{pages}{1511--1556}.
\newblock \DOIprefix\doi{10.1111/jofi.12408}.
\bibitem[{Bianchi et~al.(2020)Bianchi, Fan and
  Todorova}]{FJ-Bianchi-Fan-Todorova-2020-IntRevFinancAnal}
\bibinfo{author}{Bianchi, R.J.}, \bibinfo{author}{Fan, J.H.},
  \bibinfo{author}{Todorova, N.}, \bibinfo{year}{2020}.
\newblock \bibinfo{title}{Financialization and de-financialization of commodity
  futures: A quantile regression approach}.
\newblock \bibinfo{journal}{International Review of Financial Analysis}
  \bibinfo{volume}{68}, \bibinfo{pages}{101451}.
\newblock \DOIprefix\doi{10.1016/j.irfa.2019.101451}.
\bibitem[{Bouri et~al.(2020)Bouri, Lucey, Saeed and
  Vo}]{FJ-Bouri-Lucey-Saeed-Vo-2020-IntRevFinancAnal}
\bibinfo{author}{Bouri, E.}, \bibinfo{author}{Lucey, B.},
  \bibinfo{author}{Saeed, T.}, \bibinfo{author}{Vo, X.V.},
  \bibinfo{year}{2020}.
\newblock \bibinfo{title}{Extreme spillovers across asian-pacific currencies: A
  quantile-based analysis}.
\newblock \bibinfo{journal}{International Review of Financial Analysis}
  \bibinfo{volume}{72}, \bibinfo{pages}{101605}.
\newblock \DOIprefix\doi{10.1016/j.irfa.2020.101605}.
\bibitem[{Brennan(1958)}]{Michael-Brennan-1958-TAmericanEconomicReview}
\bibinfo{author}{Brennan, M.J.}, \bibinfo{year}{1958}.
\newblock \bibinfo{title}{The supply of storage}.
\newblock \bibinfo{journal}{American Economic Review} \bibinfo{volume}{48},
  \bibinfo{pages}{50--72}.
\bibitem[{Chen et~al.(2022a)Chen, Liang, Ding and
  Liu}]{FJ-Chen-Liang-Ding-Liu-2022-EnergyEcon}
\bibinfo{author}{Chen, J.}, \bibinfo{author}{Liang, Z.}, \bibinfo{author}{Ding,
  Q.}, \bibinfo{author}{Liu, Z.}, \bibinfo{year}{2022}a.
\newblock \bibinfo{title}{Extreme spillovers among fossil energy, clean energy,
  and metals markets: Evidence from a quantile-based analysis}.
\newblock \bibinfo{journal}{Energy Economics} \bibinfo{volume}{107},
  \bibinfo{pages}{105880}.
\newblock \DOIprefix\doi{10.1016/j.eneco.2022.105880}.
\bibitem[{Chen et~al.(2022b)Chen, Wen, Li, Yin and
  Zhao}]{FJ-Chen-Wen-Li-Yin-Zhao-2022-EnergyEcon}
\bibinfo{author}{Chen, L.}, \bibinfo{author}{Wen, F.}, \bibinfo{author}{Li,
  W.}, \bibinfo{author}{Yin, H.}, \bibinfo{author}{Zhao, L.},
  \bibinfo{year}{2022}b.
\newblock \bibinfo{title}{Extreme risk spillover of the oil, exchange rate to
  {C}hinese stock market: {E}vidence from implied volatility indexes}.
\newblock \bibinfo{journal}{Energy Economics} \bibinfo{volume}{107},
  \bibinfo{pages}{105857}.
\newblock \DOIprefix\doi{10.1016/j.eneco.2022.105857}.
\bibitem[{Cheng et~al.(2023)Cheng, Deng, Liang and
  Cao}]{FJ-Cheng-Deng-Liang-Cao-2023-ResourPolicy}
\bibinfo{author}{Cheng, S.}, \bibinfo{author}{Deng, M.},
  \bibinfo{author}{Liang, R.}, \bibinfo{author}{Cao, Y.}, \bibinfo{year}{2023}.
\newblock \bibinfo{title}{Asymmetric volatility spillover among global oil,
  gold, and {C}hinese sectors in the presence of major emergencies}.
\newblock \bibinfo{journal}{Resources Policy} \bibinfo{volume}{82},
  \bibinfo{pages}{103579}.
\newblock \DOIprefix\doi{10.1016/j.resourpol.2023.103579}.
\bibitem[{Cootner(1960)}]{Cootner-1960-JPE}
\bibinfo{author}{Cootner, P.H.}, \bibinfo{year}{1960}.
\newblock \bibinfo{title}{Returns to speculators: Telser versus keynes}.
\newblock \bibinfo{journal}{Journal of Political Economy} \bibinfo{volume}{68},
  \bibinfo{pages}{396--404}.
\newblock \DOIprefix\doi{10.1086/258347}.
\bibitem[{Diebold and Yilmaz(2009)}]{FJ-Diebold-Yilmaz-2009-EconJ}
\bibinfo{author}{Diebold, F.X.}, \bibinfo{author}{Yilmaz, K.},
  \bibinfo{year}{2009}.
\newblock \bibinfo{title}{Measuring financial asset return and volatility
  spillovers, with application to global equity markets}.
\newblock \bibinfo{journal}{Economic Journal} \bibinfo{volume}{119},
  \bibinfo{pages}{158--171}.
\newblock \DOIprefix\doi{10.1111/j.1468-0297.2008.02208.x}.
\bibitem[{Diebold and Yilmaz(2012)}]{FJ-Diebold-Yilmaz-2012-IntJForecast}
\bibinfo{author}{Diebold, F.X.}, \bibinfo{author}{Yilmaz, K.},
  \bibinfo{year}{2012}.
\newblock \bibinfo{title}{Better to give than to receive: Predictive
  directional measurement of volatility spillovers}.
\newblock \bibinfo{journal}{International Journal of Forecasting}
  \bibinfo{volume}{28}, \bibinfo{pages}{57--66}.
\newblock \DOIprefix\doi{10.1016/j.ijforecast.2011.02.006}.
\bibitem[{Diebold and Yilmaz(2014)}]{FJ-Diebold-Yilmaz-2014-JEconom}
\bibinfo{author}{Diebold, F.X.}, \bibinfo{author}{Yilmaz, K.},
  \bibinfo{year}{2014}.
\newblock \bibinfo{title}{On the network topology of variance decompositions:
  Measuring the connectedness of financial firms}.
\newblock \bibinfo{journal}{Journal of Econometrics} \bibinfo{volume}{182},
  \bibinfo{pages}{119--134}.
\newblock \DOIprefix\doi{10.1016/j.jeconom.2014.04.012}.
\bibitem[{Du and He(2015)}]{FJ-Du-He-2015-EnergyEcon}
\bibinfo{author}{Du, L.}, \bibinfo{author}{He, Y.}, \bibinfo{year}{2015}.
\newblock \bibinfo{title}{Extreme risk spillovers between crude oil and stock
  markets}.
\newblock \bibinfo{journal}{Energy Economics} \bibinfo{volume}{51},
  \bibinfo{pages}{455--465}.
\newblock \DOIprefix\doi{10.1016/j.eneco.2015.08.007}.
\bibitem[{Duan et~al.(2023)Duan, Xiao, Ren, Taghizadeh-Hesary and
  Duan}]{FJ-Duan-Xiao-Ren-TaghizadehHesary-Duan-2023-ResourPolicy}
\bibinfo{author}{Duan, X.}, \bibinfo{author}{Xiao, Y.}, \bibinfo{author}{Ren,
  X.}, \bibinfo{author}{Taghizadeh-Hesary, F.}, \bibinfo{author}{Duan, K.},
  \bibinfo{year}{2023}.
\newblock \bibinfo{title}{Dynamic spillover between traditional energy markets
  and emerging green markets: {I}mplications for sustainable development}.
\newblock \bibinfo{journal}{Resources Policy} \bibinfo{volume}{82},
  \bibinfo{pages}{103483}.
\newblock \DOIprefix\doi{10.1016/j.resourpol.2023.103483}.
\bibitem[{Gardebroek et~al.(2016)Gardebroek, Hernandez and
  Robles}]{FJ-Gardebroek-Hernandez-Robles-2016-AgricEcon}
\bibinfo{author}{Gardebroek, C.}, \bibinfo{author}{Hernandez, M.A.},
  \bibinfo{author}{Robles, M.}, \bibinfo{year}{2016}.
\newblock \bibinfo{title}{Market interdependence and volatility transmission
  among major crops}.
\newblock \bibinfo{journal}{Agricultural Economics} \bibinfo{volume}{47},
  \bibinfo{pages}{141--155}.
\newblock \DOIprefix\doi{10.1111/agec.12184}.
\bibitem[{Hernandez et~al.(2014)Hernandez, Ibarra and
  Trupkin}]{Hernandez-Ibarra-Trupkin-2014-ERAE}
\bibinfo{author}{Hernandez, M.A.}, \bibinfo{author}{Ibarra, R.},
  \bibinfo{author}{Trupkin, D.R.}, \bibinfo{year}{2014}.
\newblock \bibinfo{title}{How far do shocks move across borders? {E}xamining
  volatility transmission in major agricultural futures markets}.
\newblock \bibinfo{journal}{European Review of Agricultural Economics}
  \bibinfo{volume}{41}, \bibinfo{pages}{301--325}.
\newblock \DOIprefix\doi{10.1093/erae/jbt020}.
\bibitem[{Hung(2021)}]{FJ-Hung-2021-ResourPolicy}
\bibinfo{author}{Hung, N.T.}, \bibinfo{year}{2021}.
\newblock \bibinfo{title}{Oil prices and agricultural commodity markets:
  Evidence from pre and during {COVID}-19 outbreak}.
\newblock \bibinfo{journal}{Resources Policy} \bibinfo{volume}{73},
  \bibinfo{pages}{102236}.
\newblock \DOIprefix\doi{10.1016/j.resourpol.2021.102236}.
\bibitem[{Iqbal et~al.(2023)Iqbal, Bouri, Grebinevych and
  Roubaud}]{FJ-Iqbal-Bouri-Grebinevych-Roubaud-MISSING-AnnOperRes}
\bibinfo{author}{Iqbal, N.}, \bibinfo{author}{Bouri, E.},
  \bibinfo{author}{Grebinevych, O.}, \bibinfo{author}{Roubaud, D.},
  \bibinfo{year}{2023}.
\newblock \bibinfo{title}{Modelling extreme risk spillovers in the commodity
  markets around crisis periods including {COVID}19}.
\newblock \bibinfo{journal}{Annals of Operations Research}
  \DOIprefix\doi{10.1007/s10479-022-04522-9}.
\bibitem[{Ji et~al.(2018)Ji, Bouri, Roubaud and
  Shahzad}]{FJ-Ji-Bouri-Roubaud-Shahzad-2018-EnergyEcon}
\bibinfo{author}{Ji, Q.}, \bibinfo{author}{Bouri, E.},
  \bibinfo{author}{Roubaud, D.}, \bibinfo{author}{Shahzad, S.J.H.},
  \bibinfo{year}{2018}.
\newblock \bibinfo{title}{Risk spillover between energy and agricultural
  commodity markets: A dependence-switching {CoVaR}-copula model}.
\newblock \bibinfo{journal}{Energy Economics} \bibinfo{volume}{75},
  \bibinfo{pages}{14--27}.
\newblock \DOIprefix\doi{10.1016/j.eneco.2018.08.015}.
\bibitem[{Jia et~al.(2016)Jia, Wang, Tu and
  Li}]{FJ-Jia-Wang-Tu-Li-2016-PhysicaA}
\bibinfo{author}{Jia, R.L.}, \bibinfo{author}{Wang, D.H.}, \bibinfo{author}{Tu,
  J.Q.}, \bibinfo{author}{Li, S.P.}, \bibinfo{year}{2016}.
\newblock \bibinfo{title}{Correlation between agricultural markets in dynamic
  perspective-evidence from {C}hina and the {US} futures markets}.
\newblock \bibinfo{journal}{Physica A} \bibinfo{volume}{464},
  \bibinfo{pages}{83--92}.
\newblock \DOIprefix\doi{10.1016/j.physa.2016.07.048}.
\bibitem[{Jiang et~al.(2016)Jiang, Su, Todorova and
  Roca}]{FJ-Jiang-Su-Todorova-Roca-2016-JFuturesMark}
\bibinfo{author}{Jiang, H.}, \bibinfo{author}{Su, J.J.},
  \bibinfo{author}{Todorova, N.}, \bibinfo{author}{Roca, E.},
  \bibinfo{year}{2016}.
\newblock \bibinfo{title}{Spillovers and directional predictability with a
  cross-quantilogram analysis: The case of {US} and {C}hinese agricultural
  futures}.
\newblock \bibinfo{journal}{Journal of Futures Markets} \bibinfo{volume}{36},
  \bibinfo{pages}{1231--1255}.
\newblock \DOIprefix\doi{10.1002/fut.21779}.
\bibitem[{Kang et~al.(2017)Kang, McIver and
  Yoon}]{FJ-Kang-McIver-Yoon-2017-EnergyEcon}
\bibinfo{author}{Kang, S.H.}, \bibinfo{author}{McIver, R.},
  \bibinfo{author}{Yoon, S.M.}, \bibinfo{year}{2017}.
\newblock \bibinfo{title}{Dynamic spillover effects among crude oil, precious
  metal, and agricultural commodity futures markets}.
\newblock \bibinfo{journal}{Energy Economics} \bibinfo{volume}{62},
  \bibinfo{pages}{19--32}.
\newblock \DOIprefix\doi{10.1016/j.eneco.2016.12.011}.
\bibitem[{Ke et~al.(2019)Ke, Li, McKenzie and
  Liu}]{FJ-Ke-Li-McKenzie-Liu-2019-Sustainability}
\bibinfo{author}{Ke, Y.}, \bibinfo{author}{Li, C.}, \bibinfo{author}{McKenzie,
  A.M.}, \bibinfo{author}{Liu, P.}, \bibinfo{year}{2019}.
\newblock \bibinfo{title}{Risk transmission between {C}hinese and {US}
  agricultural commodity futures markets - a {CoVaR} approach}.
\newblock \bibinfo{journal}{Sustainability} \bibinfo{volume}{11},
  \bibinfo{pages}{239}.
\newblock \DOIprefix\doi{10.3390/su11010239}.
\bibitem[{Koekebakker and Lien(2004)}]{FJ-Koekebakker-Lien-2004-AmJAgrEcon}
\bibinfo{author}{Koekebakker, S.}, \bibinfo{author}{Lien, G.},
  \bibinfo{year}{2004}.
\newblock \bibinfo{title}{Volatility and price jumps in agricultural future
  prices - {E}vidence from wheat options}.
\newblock \bibinfo{journal}{American Journal of Agricultural Economics}
  \bibinfo{volume}{86}, \bibinfo{pages}{1018--1031}.
\newblock \DOIprefix\doi{10.1111/j.0002-9092.2004.00650.x}.
\bibitem[{Koenker and Bassett(1978)}]{FJ-Koenker-Roger-1978-Econometrica}
\bibinfo{author}{Koenker, R.}, \bibinfo{author}{Bassett, G.},
  \bibinfo{year}{1978}.
\newblock \bibinfo{title}{Regression quantiles}.
\newblock \bibinfo{journal}{Econometrica} \bibinfo{volume}{46},
  \bibinfo{pages}{33–50}.
\newblock \DOIprefix\doi{10.2307/1913643}.
\bibitem[{Kristoufek et~al.(2012)Kristoufek, Janda and
  Zilberman}]{FJ-Kristoufek-Janda-Zilberman-2012-EnergyEcon}
\bibinfo{author}{Kristoufek, L.}, \bibinfo{author}{Janda, K.},
  \bibinfo{author}{Zilberman, D.}, \bibinfo{year}{2012}.
\newblock \bibinfo{title}{Correlations between biofuels and related commodities
  before and during the food crisis: A taxonomy perspective}.
\newblock \bibinfo{journal}{Energy Economics} \bibinfo{volume}{34},
  \bibinfo{pages}{1380--1391}.
\newblock \DOIprefix\doi{10.1016/j.eneco.2012.06.016}.
\bibitem[{Li and Xiong(2021)}]{FJ-Li-Xiong-2021-JAsianEcon}
\bibinfo{author}{Li, M.}, \bibinfo{author}{Xiong, T.}, \bibinfo{year}{2021}.
\newblock \bibinfo{title}{Dynamic price discovery in {C}hinese agricultural
  futures markets}.
\newblock \bibinfo{journal}{Journal of Asian Economics} \bibinfo{volume}{76},
  \bibinfo{pages}{101370}.
\newblock \DOIprefix\doi{10.1016/j.asieco.2021.101370}.
\bibitem[{Luo and Ji(2018)}]{FJ-Luo-Ji-2018-EnergyEcon}
\bibinfo{author}{Luo, J.}, \bibinfo{author}{Ji, Q.}, \bibinfo{year}{2018}.
\newblock \bibinfo{title}{High-frequency volatility connectedness between the
  {US} crude oil market and {C}hina's agricultural commodity markets}.
\newblock \bibinfo{journal}{Energy Economics} \bibinfo{volume}{76},
  \bibinfo{pages}{424--438}.
\newblock \DOIprefix\doi{10.1016/j.eneco.2018.10.031}.
\bibitem[{McKenzie and Holt(2002)}]{FJ-McKenzie-Holt-2002-ApplEcon}
\bibinfo{author}{McKenzie, A.}, \bibinfo{author}{Holt, M.},
  \bibinfo{year}{2002}.
\newblock \bibinfo{title}{Market efficiency in agricultural futures markets}.
\newblock \bibinfo{journal}{Applied Economics} \bibinfo{volume}{34},
  \bibinfo{pages}{1519--1532}.
\newblock \DOIprefix\doi{10.1080/00036840110102761}.
\bibitem[{Mensi et~al.(2014)Mensi, Hammoudeh, Nguyen and
  Yoon}]{FJ-Mensi-Hammoudeh-Nguyen-Yoon-2014-EnergyEcon}
\bibinfo{author}{Mensi, W.}, \bibinfo{author}{Hammoudeh, S.},
  \bibinfo{author}{Nguyen, D.K.}, \bibinfo{author}{Yoon, S.M.},
  \bibinfo{year}{2014}.
\newblock \bibinfo{title}{Dynamic spillovers among major energy and cereal
  commodity prices}.
\newblock \bibinfo{journal}{Energy Economics} \bibinfo{volume}{43},
  \bibinfo{pages}{225--243}.
\newblock \DOIprefix\doi{10.1016/j.eneco.2014.03.004}.
\bibitem[{Su(2020)}]{FJ-Su-2020-NAmEconFinanc}
\bibinfo{author}{Su, X.}, \bibinfo{year}{2020}.
\newblock \bibinfo{title}{Measuring extreme risk spillovers across
  international stock markets: A quantile variance decomposition analysis}.
\newblock \bibinfo{journal}{North American Journal of Economics and Finance}
  \bibinfo{volume}{51}, \bibinfo{pages}{101098}.
\newblock \DOIprefix\doi{10.1016/j.najef.2019.101098}.
\bibitem[{Tiwari et~al.(2022)Tiwari, Abakah, Dwumfour and
  Mefteh-Wali}]{FJ-Tiwari-Abakah-Dwumfour-MeftehWali-2022-ApplEcon}
\bibinfo{author}{Tiwari, A.K.}, \bibinfo{author}{Abakah, E.J.A.},
  \bibinfo{author}{Dwumfour, R.A.}, \bibinfo{author}{Mefteh-Wali, S.},
  \bibinfo{year}{2022}.
\newblock \bibinfo{title}{Connectedness and directional spillovers in energy
  sectors: International evidence}.
\newblock \bibinfo{journal}{Applied Economics} \bibinfo{volume}{54},
  \bibinfo{pages}{2554--2569}.
\newblock \DOIprefix\doi{10.1080/00036846.2021.1998326}.
\bibitem[{Tiwari et~al.(2021)Tiwari, Boachie, Suleman and
  Gupta}]{FJ-Tiwari-Boachie-Suleman-Gupta-2021-Energy}
\bibinfo{author}{Tiwari, A.K.}, \bibinfo{author}{Boachie, M.K.},
  \bibinfo{author}{Suleman, M.T.}, \bibinfo{author}{Gupta, R.},
  \bibinfo{year}{2021}.
\newblock \bibinfo{title}{Structure dependence between oil and agricultural
  commodities returns: The role of geopolitical risks}.
\newblock \bibinfo{journal}{Energy} \bibinfo{volume}{219},
  \bibinfo{pages}{119584}.
\newblock \DOIprefix\doi{10.1016/j.energy.2020.119584}.
\bibitem[{Yip et~al.(2020)Yip, Brooks, Do and
  Nguyen}]{FJ-Yip-Brooks-Do-Nguyen-2020-IntRevFinancAnal}
\bibinfo{author}{Yip, P.S.}, \bibinfo{author}{Brooks, R.}, \bibinfo{author}{Do,
  H.X.}, \bibinfo{author}{Nguyen, D.K.}, \bibinfo{year}{2020}.
\newblock \bibinfo{title}{Dynamic volatility spillover effects between oil and
  agricultural products}.
\newblock \bibinfo{journal}{International Review of Financial Analysis}
  \bibinfo{volume}{69}, \bibinfo{pages}{101465}.
\newblock \DOIprefix\doi{10.1016/j.irfa.2020.101465}.
\bibitem[{Zivkov et~al.(2020)Zivkov, Kuzman and
  Subic}]{FJ-Zivkov-Kuzman-Subic-2020-AgricEcon}
\bibinfo{author}{Zivkov, D.}, \bibinfo{author}{Kuzman, B.},
  \bibinfo{author}{Subic, J.}, \bibinfo{year}{2020}.
\newblock \bibinfo{title}{What {B}ayesian quantiles can tell about volatility
  transmission between the major agricultural futures?}
\newblock \bibinfo{journal}{Agricultural Economics-Zemedelska Ekonomika}
  \bibinfo{volume}{66}, \bibinfo{pages}{215--225}.
\newblock \DOIprefix\doi{10.17221/127/2019-AGRICECON}.

\end{thebibliography}

\end{document}